\documentclass[aps,prl,twocolumn,showpacs,superscriptaddress,groupedaddress,floatfix]{revtex4}  % 

\usepackage{rotating}

\usepackage{natbib}
\usepackage{amsmath,amssymb}% Include figure files
\usepackage{graphicx}% Include figure files
\usepackage{dcolumn}% Align table columns on decimal point
\usepackage{bm}% bold math

\newcommand{\beq}{\begin{equation}}
\newcommand{\eeq}{\end{equation}}

\newcommand{\fpv}{f_p({\mathbf{v}})}

\begin{document}

\title{The In Situ Signature of Cyclotron Resonant Heating}

\author{Trevor A. Bowen}\email{tbowen@berkeley.edu}
\affiliation{Space Sciences Laboratory, University of California, Berkeley, CA 94720-7450, USA}

\author{Benjamin D.~G. {Chandran}}
\affiliation{Department of Physics \& Astronomy, University of New Hampshire, Durham, NH 03824, USA}
\author{Jonathan Squire}
\affiliation{Department of Physics, University of Otago, 730 Cumberland St., Dunedin 9016, New
Zealand}
\author{Stuart D. Bale}
\affiliation{Space Sciences Laboratory, University of California, Berkeley, CA 94720-7450, USA}
\affiliation{Physics Department, University of California, Berkeley, CA 94720-7300, USA}

\author{Die Duan}
\affiliation{School of Earth and Space Sciences, Peking University, Beijing, 100871, China}

\author{Kristopher G. Klein}
\affiliation{Department of Planetary Sciences \& Lunar and Planetary Laboratory, University of Arizona, Tucson, AZ 85721}
\author{Davin Larson}
\affiliation{Space Sciences Laboratory, University of California, Berkeley, CA 94720-7450, USA}
\author{Alfred Mallet}
\affiliation{Space Sciences Laboratory, University of California, Berkeley, CA 94720-7450, USA}
\author{Michael D. McManus}
\affiliation{Space Sciences Laboratory, University of California, Berkeley, CA 94720-7450, USA}
\affiliation{Physics Department, University of California, Berkeley, CA 94720-7300, USA}

\author{Romain Meyrand}
\affiliation{Department of Physics, University of Otago, 730 Cumberland St., Dunedin 9016, New
Zealand}
\author{Jaye L. Verniero}
\affiliation{NASA Goddard Space Flight Center}
\author{Lloyd D. Woodham}
\affiliation{Department of Physics, The Blackett Laboratory,Imperial College London, London, SW7 2AZ, UK}

\begin{abstract}
   The dissipation of magnetized turbulence is an important paradigm for describing heating and energy transfer in astrophysical environments such as the solar corona and wind; however, the specific collisionless processes behind dissipation and heating remain relatively unconstrained by measurements. Remote sensing observations have suggested the presence of strong temperature anisotropy in the solar corona consistent with cyclotron resonant heating. In the solar wind, in situ magnetic field measurements reveal the presence of cyclotron waves, while measured ion velocity distribution functions have hinted at the active presence of cyclotron resonance. Here, we present Parker Solar Probe observations that connect the presence of ion-cyclotron waves directly to signatures of resonant damping in observed proton-velocity distributions using the framework of quasilinear theory. We show that the quasilinear evolution of the observed distribution functions should absorb the observed cyclotron wave population with a heating rate of $10^{-14}$ W/m$^3$, indicating significant heating of the solar wind.\end{abstract}
\maketitle %remain largely elusive. The 
\paragraph{Introduction} 
Observations of the solar corona reveal plasma that is millions of degrees hotter than the blackbody temperature of the solar surface. While the energy required to heat the corona, and accelerate the solar wind originates from solar convection and the magnetic fields produced by the solar dynamo, the specific pathways to heating and particle acceleration remain elusive \cite{Marsch2006}.  The dissipation of Alfv\'{e}nic turbulence at kinetic scales has become a common paradigm in explaining the dynamics of coronal heating and solar wind acceleration \cite{Coleman1968,MatthaeusGoldstein1982,CranmervanBallegooijen2003,Chandran2010a}; % Though the specific mechanisms governing dissipation remain under debate, inter-particle collisions in the solar wind and corona are infrequent, implying that these processes must be collisionless in nature \cite{Kasper2008}.
possible dissipative mechanisms include Landau or cyclotron resonant damping \cite{Barnes1966, Barnes1979,Goldstein1994,Leamon1998b, Howes2011}, stochastic heating  \cite{Chandran2013}, or magnetic reconnection \cite{Mallet2017,Loureiro2017a}. %It is additionally unclear whether turbulent dissipation deposits energy directly into ions directly, or if secondary processes related to kinetic \cite{Bale2009,Alexandrova2013,Klein2018} and fluid \cite{Hellinger2006,Matteini2007,Bowen2018,Tenerani2020} instabilities, or  intermittent effects and coherent structures \citep{Alexandrova2008b,Osman2011,Lion2016,Mallet2019,Chhiber2021,Sioulas2022} play a significant role. 
Additionally, the portion of energy deposited by these processes at ion scales, versus that which is subject to a kinetic cascade and dissipated by electrons, remains an open question \cite{QuataertGruzinov1999,Chandran2010b,Howes2011,Chandran2011,Alexandrova2012}.
%Due to the small scales associated with the dissipation, sensitive in-situ measurements at high time resolution are needed to distinguish between various theoretical mechanisms for turbulent dissipation and plasma heating. Signatures of heating are apparent in the ion velocity distribution function $\fv$ \cite{Marsch2006}. 

It is well known that the observed ion temperature profiles in the solar wind require significant perpendicular heating \citep{Richardson1995,Matteini2007}, which is likely initiated at ion kinetic scales where particles interact efficiently with electromagnetic waves \citep{Quataert1998,Leamon1998a,Leamon1998b,Gary1999,Leamon2000,Howes2008,Chandran2011}. %Such heating transfers energy from the waves to the particles, generating observable electromagnetic signatures and features in the distribution function \cite{MarschTu2001,He2015,Verniero2020}.
%M%easurements of 3D ion distributions will enable comparison of damping rates associated with, e.g. Landau and cyclotron resonance, with our observed dissipation rates. \citep{Lysak1996,Leamon1998b,Quataert1998,QuataertGruzinov1999,Leamon1999,Gary1999,Leamon2000,Howes2008,Sahraoui2010,Chandran2011,Smith2012,Verniero2020}. Additionally, heating processes may leave clear signatures in the distribution function: e.g. diffusive shells in the case of cyclotron resonance \cite{Isenberg2004}; or flattening associated with resonant damping or perpendicular stochastic heating \cite{He2015,Martinovic2020}. 
Cyclotron resonant coupling of electromagnetic fluctuations with ion gyromotion  \cite{Stix1992}, has received particular attention as a potential perpendicular heating mechanism \citep{HollwegJohnson1988,TuMarsch1997,Cranmer2000,HollwegIsenberg2002,Cranmer2014}.
Ultraviolet spectroscopic measurements of coronal ion temperature anisotropy suggest large $T_\perp/T_\parallel$, consistent with cyclotron resonant heating \citep{Kohl1997,Kohl1998,Cranmer1999,Cranmer2000}. 
The presence of ion-cyclotron waves has been well documented in in-situ observations throughout the heliosphere both as solitary waves and as part of the background spectrum of fluctuations \cite{Jian2009,Podesta2011,He2011,Wicks2016,Woodham2019,Bale2019,Bowen2020a}. Observations of magnetic-helicity at ion scales have been interpreted as evidence for active cyclotron damping of quasi-parallel Alfv\'{e}nic fluctuations, which contribute to turbulent heating \citep{Isenberg1990,Goldstein1994,Leamon1998b,Woodham2018}. 

Theoretical signatures of resonant interactions in particle distribution functions are often studied in the framework of quasilinear (QL) diffusion \cite{KennelEngelmann1966,IsenbergLee1996,Isenberg2001,Chandran2010b}; observations of the solar wind have suggested evidence for QL cyclotron resonant diffusion in signatures of the proton velocity distribution function $\fpv$ \cite{MarschTu2001,MarschBourouaine2011,He2015}. While the generation of cyclotron waves through instabilities has been widely discussed \cite{Gary2001,Gary2016,Wicks2016,Woodham2019,Bowen2020a} and signatures of cyclotron resonant dissipation have been suggested \cite{Goldstein1994,Leamon1998b,MarschTu2001,Kasper2008,He2015,Telloni2019,Vech2021}, definitive cyclotron resonant heating sufficient to power the solar wind has not been observed.
%While the presence of ion-cyclotron waves has been well documented throughout the heliosphere both as solitary waves and as part of the background spectrum of fluctuations \cite{Jian2009,Podesta2011,He2011,Wicks2016,Woodham2019,Bale2019,Bowen2020a}. 
%In addition to proton-cyclotron resonance, the cyclotron resonance of doubly ionized helium ($\alpha$-particles) and heavy ions has been studied extensively \citep{IsenbergHollweg1983,Isenberg1984,Kasper2008}. Using observations of protons and $\alpha$-particles \cite{Kasper2013} argue that temperature anisotropy observed at 1AU is consistent with {\em{in siitu}} cyclotron resonant heating; though alternative theories may produce consistent solutions with other dissipation mechanisms, e.g. stochastic heating \citep{Chandran2010a,Chandran2013}.

%The signature of cyclotron damping on distribution functions has been studied thoroughly in the framework of QL diffusion\cite{KennelEngelmann1966,IsenbergLee1996,MarschTu2001,Isenberg2001,Chandran2010b}. Observations from Helios and Wind have suggested that $\fpv$ may be shaped by QL diffusion of cyclotron resonant particles \cite{MarschTu2001,MarschBourouaine2011,He2015}. Recent work has additionally demonstrated resonance of right handed modes with broadened distribution functions in the inner helsiophere \cite{Verniero2021}.

\begin{figure}[!]
    \centering
    \includegraphics[width=3.5in]{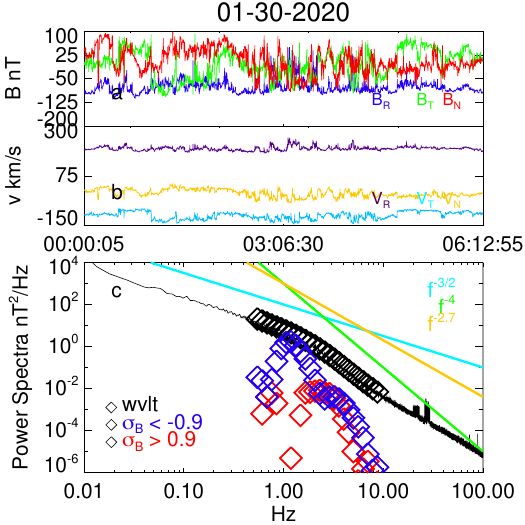}
   \caption{(a) Magnetic field measurements from PSP/FIELDS. (b) Velocity measurements from PSP/SPANi. (c) Spectra of magnetic field data. Spectral indices at -3/2,-4 and -2.7 are shown. Wavelet coefficients for total power are shown as black $\diamond$. Wavelet coefficients filtered by left/right handed $\sigma_b$ is shown in red/blue $\diamond$.}
    \label{fig:1}
\end{figure}

 In this Letter, we apply the QL theory of resonant cyclotron interactions \cite{KennelEngelmann1966,IsenbergLee1996} to empirically measured cyclotron wave spectra and ion-distribution functions. Our results provide evidence of substantial heating at levels comparable with bulk solar wind heating rates, providing a compelling picture of ion-heating in the solar wind.

\paragraph{Data}
Parker Solar Probe (PSP, \cite{Fox2016}) observations from the electromagnetic FIELDS \citep{Bale2016} and Solar Wind Electron Alpha and Proton (SWEAP, \cite{Kasper2016}) instruments aim to constrain fundamental processes of around coronal heating and solar wind acceleration. PSP has revealed prevalent ion-scale electromagnetic waves \cite{Bale2019,Bowen2020a}, ion-distributions out of thermal equilibrium \cite{Verniero2020,Klein2021}, and evidence for resonant wave-particle interactions predicted by QL theory \cite{Verniero2021}. To constrain cyclotron resonant heating, we study a stream from PSP perihelion 4 from 2020-01-30/00:00-08:00 with resolved proton distributions. During the interval PSP  was $\sim 30 R_\odot$ from the solar surface. We use merged search coil and fluxgate magnetometer data from PSP/FIELDS \cite{Bale2016,Bowen2020b} enabling measurement of the inertial, transition, and kinetic scales of turbulence; the merged data set only has two axes available \cite{DudokdeWit2022}, thus we use vector-fluxgate magnetometer data to study wave-polarization. Fig \ref{fig:1}(a) shows  $\mathbf{B}$ in Radial-Tangential-Normal (RTN) coordinates. Proton velocity distribution functions $\fpv$ are obtained from the  PSP/SWEAP Solar Probe ANalyzer (SPANi). The proton population is often parameterized with a pair of drifting bi-Maxwellian fits to model $\fpv$ using separate thermal (core) and non-thermal (beam) populations  \cite{Marsch2006}. Fits to a proton core and field-aligned beam provide estimates of bulk velocity ${\bf{u}}$, anisotropic temperatures perpendicular and parallel the background magnetic field $T_{\parallel,\perp}$, and the beam-to-core proton density ratio $n_b/n_c$ \cite{Kasper2016,Klein2021}. 
Fig \ref{fig:1}(b) shows measurements of ${\bf{u}}$ in RTN coordinates. The stream is relatively slow with an average speed of $\sim $220 km/s, and moderately Alfv\'{e}nic with a cross helicity of $\sim$0.85.
 
The phase-space density of $\fpv$ is calibrated to quasi-thermal noise from FIELDS to recover the absolute density \cite{Bale2016,Pulupa2017}. The mean proton density is $n_p=1100/$cm$^3$, SPANi gives an average beam to core density ratio of 0.48. The core/beam have $T_{\perp}$ of 15 eV/22 eV and $T_{\parallel}$ of 12 eV/30 eV. The average drift of the beam relative to the core is 83 km/s. The individual core and beam have $\beta_{c}=0.65$ and  $\beta_{b}=1.1$. The mean magnetic field was directed sunward, with an Alfv\'{e}n speed of 60 km/s.
 Fig \ref{fig:1}(c) shows the magnetic field spectra of the interval with a steep transition range at ion-kinetic scales \cite{Sahraoui2009,Kiyani2009,Bowen2020c}. %The transition range slope of $-4$ is associated with ion-scale turbulent dissipation \cite{Bowen2020c}.
 %A single component bi-Maxwellian proton population with equivalent macroscopic thermodynamic properties to the drifting bi-Maxwellian fit has parallel and perpendicular temperatures of 29 eV and 18 eV, with $T_\perp/T_\parallel < 1$ and $\beta=0.96$ \cite{Klein2021}. 

%\paragraph{Cyclotron Waves}% We consider frequencies (assuming Taylor's hypothesis) corresponding to the ion gyroscale and cyclotron resonance of the thermal speed \cite{Leamon1998a}
 
 %\begin{align}
   % f_\rho&=\frac{v_{sw}}{2\pi}\frac{qB_0}{m_pv_{\perp pth}}\\
  %  f_\Omega&=\frac{v_{sw}}{2\pi}\frac{\Omega_p}{v_A +v_{\parallel pth}}.
  % \end{align}
  
  %\begin{align}
  %%  f_\Omega&=\frac{k_\Omega v_{\parallel sw} +\omega}{2\pi}%=\frac{v_{sw}}{2\pi}\frac{\Omega_p}{V_A +v_{\parallel pth}},
 %   \end{\align}
  %\end{align}
%where $k_\Omega$ is the resonant wave-number of the parallel thermal speed and $\omega\$.
We apply a Morlet wavelet transform to the vector magnetic field data rotated into field aligned coordinates \cite{TorrenceCompo1998}.  %$\hat{W}=(\hat{B}_{\perp1},\hat{B}_{\perp2},\hat{B}_0)$.  %\begin{equation}
 %  W(s,t)=\sum_{i=0}^{N-1} \%psi\left(\frac{t_i-\tau}{s}\right)B(t_i).\\
  % \end{equation}
  Signatures of circular polarization are found using
\begin{equation}
        \sigma_B(f,t)=-2\text{Im}(B_{\perp1}B_{\perp2}^*)/(B_{\perp1}^2+B_{\perp2}^2),
\end{equation}
 with left/right handed waves corresponding to positive/negative $\sigma_B$ \cite{HowesQuataert2010,Podesta2011,He2011,Klein2014}.
 Circular polarization is measured in the spacecraft frame, such that the measured sign may not correspond to the innate plasma frame polarization \cite{HowesQuataert2010}. A sign change in $\sigma_B$ occurs if the wave is Doppler shifted to negative frequencies in the spacecraft frame. However, it has been demonstrated that the majority of waves propagate outward, and thus, that Doppler shift does not change their handedness when observed in the spacecraft frame \cite{Bowen2020d}. %Recent results suggest that the fast magnetosonic instability may be the dominant instability in distributions with significant proton beam densities \cite{Verniero2020,Klein2021}.

 Previous work has shown that circularly polarized ion-scale waves are parallel propagating and evident when $\theta_{vB}\sim 0$ \cite{Bowen2020a}. However, observations of parallel propagating, circular polarized, waves are strongly inhibited when the angle between the solar wind and the mean magnetic field is oblique. This effect occurs because a) the wave polarization plane is not well resolved by the spacecraft and b) the turbulence is anisotropic with increasing power with larger  $\theta_{vB}$ \cite{Fredricks1976,Horbury2008,HowesQuataert2010,Horbury2012}. The lack of circular polarization signatures when $\theta_{vB}$ is moderately oblique is consistent with sampling effects of quasi-parallel waves at oblique angles in anisotropic turbulence \cite{Bowen2020a}, suggesting that ion-scale waves can persist at oblique $\theta_{vB}$. In order to estimate the parallel propagating left-hand polarized spectrum, wavelet power with $\sigma_B >0.9$ is identified when $\theta_{vB} <15^{\circ}$. We assume homogeneity, and stationarity, such that the circularly polarized spectrum measured at $\theta_{vB}<15^{\circ}$ represents the wave spectrum at all times (i.e. when $\theta_{vB}>15^{\circ}$). Fig 1(c) shows the power spectrum of circularly polarized fluctuations with $\sigma_B >0.9$ and $\sigma_B< -0.9$, corresponding to strong left and right-handed power. The right handed modes have been shown to be statistically consistent with a fast magnetosonic mode \cite{Bowen2020d}.

\begin{figure}[!]
    \centering
    \includegraphics[width=3.5in]{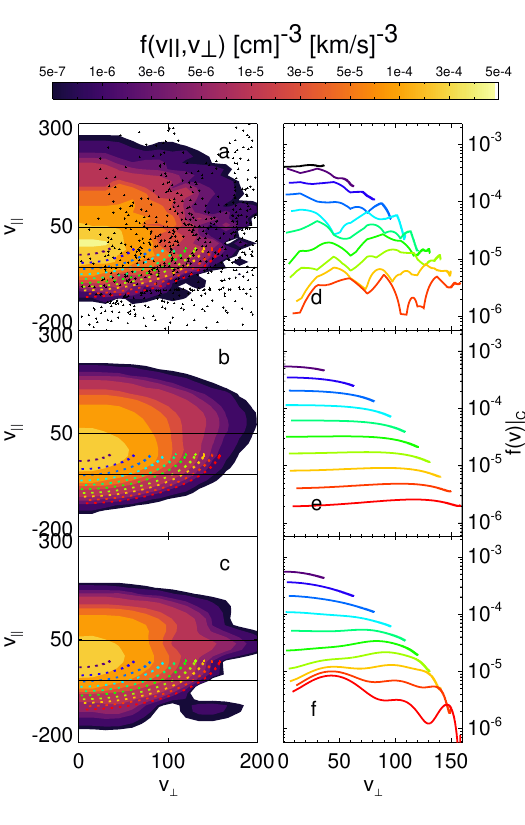}
   \caption{(a) Interpolation of $\fpv$ from SPANi in the $v_\perp-v_\parallel$ plane, with the mean magnetic field pointing vertically. Points show SPANi measurement locations. Solid lines show $v_{\parallel{th}}$, and set of cyclotron-resonant diffusion contours are plotted. (b) Drifting bi-Maxwellian fit to $\fpv$. (c) Hermite representation of $\fpv$.  Integration of $\fpv$ in a-c over $v_\perp-v_\parallel$ is normalized to the QTN density. (d) Contours of $\fpv$ determined by interpolating the gyrotropic distribution along QL diffusion contours for parallel cyclotron resonance. (e-f) Drifting bi-Maxwellian and the Hermite representations of $\fpv$ evaluated along cyclotron resonance contours.}
    \label{fig:2}
\end{figure}
     
 \paragraph{Distribution Functions}  Fig 2 shows the SPANi proton distribution function $f_p(v_\perp, v_\parallel)$ from the stream at 2020-01-30/04:10:21. Figure 2(a) shows a interpolation of the 3D measurements in the $v_\perp-v_\parallel$ plane constructed by identifying values of $v_\perp$ and $v_\parallel$ for each 3D energy bin, assuming gyrotropy. A drifting two-component bi-Maxwellian fit, assuming gyrotropy, is shown in Fig 2(b). The drifting bi-Maxwellian fit provides an approximation to $\fpv$ using two individual proton populations, though this parametrization may not resolve all non-thermal features that affect resonant interaction with cyclotron waves. Indeed, the presence of strong and persistent  cyclotron-resonant interactions should affect the shape of $\fpv$, leading to an equilibrium distribution that deviates from a bi-Maxwellian \cite{IsenbergLee1996}. To explore a nonparametric $f_p(v_\perp, v_\parallel)$, which may better represent the data \cite{Dum1980,Vinas2009,Servidio2017}, we fit a set of orthonormal Hermite functions using linear least square methods:
\begin{align} 
f_p(v_\perp, v_\parallel)= \sum_{m,n} g_{mn}\phi_m(v_\perp/v_{\perp{th}})\phi_n(v_\parallel/v_{\parallel{th}})\\
H_n(v) = (-1)^n e^{v^2}\frac{d^n}{dx^n}e^{-v^2}\\
\phi_m=\frac{H_m(v)}{\sqrt{2^m \pi^{1/2} m!}}e^{-v^2}
%\phi^m=\frac{H_m(v)}{\sqrt{2^m \pi^{1/2} m!}}
%\int\phi_m\phi^n dv=\int\phi^m\phi^n e^{-v^2} dv=\delta_{n,m}
\end{align}
\cite{Parker2015,Schekochihin2016}.
  Fig 2(c) shows the best-fit estimate to $\fpv$ for Hermite functions of order $m_{max}=6, n_{max}=6$. The distributions are shown in field aligned coordinates with $\hat{\mathbf{B}}_0$ along the vertical axis. In carrying out this fit, we effectively extend $f(v_\perp, v_\parallel)$ to negative values of $v_\perp$ by treating $f(v_\perp, v_\parallel)$ as an even function of $v_{\perp}$, thereby omitting the terms in the sum corresponding to odd values of $m$. Our use of Hermite functions is meant solely as a nonparametric interpolative scheme, and is not intended to represent a natural-basis for $\fpv$ \cite{Parker2015}. Over the studied interval, the average $\chi^2$ residual of the Hermite representation is 90\% of the bi-Maxwellian fit. No intervals were significantly better fit by the drifting bi-Maxwellian, though some distributions are equally well represented by either approximation.
 
The ion cyclotron resonance condition is  
$\omega(k_\parallel) =\Omega +k_\parallel v_\parallel$
such that outward-propagating cyclotron waves resonate with the inward propagating portion of the distribution function. The evolution of $\fpv$ in the presence of resonant interactions can be described by QL diffusion theory \cite{KennelEngelmann1966,IsenbergLee1996}. In a reference frame moving with a wave, the particles conserve kinetic energy as they scatter off that wave, tracing contours in $v_\perp$ and $v_\parallel$ that can be computed using the wave dispersion relation and resonance condition \cite{KennelEngelmann1966,IsenbergLee1996,MarschTu2001,HollwegIsenberg2002}. %Defining  $y={k_\parallel v_A}/{\Omega}$, the diffusion contours for parallel cold-plasma ion cyclotron waves \cite{IsenbergLee1996,Squire2021},
   %  \begin{align}
   %  \frac{v_\perp^2}{v_a^2} +\frac{1}{y^2} - \text{sinh}^{-1}y/2 +\text{ln} y=const,\end{align}  with $y(v_\parallel)$ defined implicitly through Equation \ref{eq:icw res}, 
   The QL diffusion contours \cite{IsenbergLee1996,Squire2020} are overlaid on $\fpv$ in Fig 2(a-c). If $\fpv$ decreases as $v_\perp$ increases along the contours, cyclotron resonance diffuses energy across in the region where resonant waves are present, heating the plasma. Conversely, if $\fpv$ increases as $v_\perp$ increases, then $\fpv$ is unstable, generating waves and cooling the plasma. Cyclotron resonant equilibrium corresponds to a flattening of $\fpv$ along the contours. Fig 2(d-f) shows $\fpv$ evaluated along QL diffusion contours, parameterized on $v_\perp$; for each representation, $\fpv$ is characteristically flat along contours, suggesting that $\fpv$ has been processed by cyclotron-resonance \cite{MarschTu2001,He2015}. %While flattening along the contours is an important signature of cyclotron resonance, a consideration of heating rates is necessary to determine whether plasma is cooled or heated  by the wave-particle interaction.
     
     %.  Figure 2(d) shows the bi-Maxwellian $\fpv$ evaluated on the quasiliner diffusion contours. 

     %The diffusion contours in Figure 2(i) are defined by solutions to the QL diffusion operator and take the form of
     %%\frac{v_\perp^2}{v_a^2} +\frac{1}{y^2} - \text{sinh}^{-1}y/2 +\text{ln} %y=const\\
    % y=\frac{k_\parallel v_A}{\Omega}\end{align}
  
  The QL proton heating rate due to resonance with parallel-propagating cyclotron waves is given by

\begin{equation}\label{eq:H}
\begin{split}
&\mathcal{H}=\int \frac{m_pv^2}{2}\frac{\partial\fpv}{\partial t} d^3\mathbf{v} =\\&\frac{\pi e^2}{2m_p^2}\int_0^\infty dk_\parallel \frac{1}{v_\perp} \hat{G}_k v_\perp \delta(\omega_k -k_\parallel v_\parallel-\Omega_p)\frac{\omega_k^2}{k_\parallel^2c^2} I(k_\parallel)\hat{G}_k \fpv d^3\mathbf{v} 
\end{split}
\end{equation}
%    \begin{align} \mathcal{H}=\int \frac{m_pv^2}{2}\frac{d\fpv}{dt} d^3\mathbf{v} &=\\\frac{\pi e^2}{2m_p^2}\int_0^\infty dk_\parallel \frac{1}{v_\perp} \hat{G}_k v_\perp \delta(\omega_k -k_\parallel v_\parallel-\Omega_p)\frac{\omega_k^2}{k_\parallel^2c^2} I(k_\parallel)\hat{G} \fpv
%    \end{align}
    with
    \begin{align}
    \hat{G}_k=  (1- \frac{k_\parallel v_\parallel}{\omega})\frac{\partial}{\partial{v_\perp}} +\frac{k_\parallel v_\perp}{\omega}\frac{\partial}{\partial{v_\parallel}}\end{align} 
\cite{KennelEngelmann1966}. Using the observed left-handed spectrum in Fig 1, the cold-plasma dispersion relation, along with the bulk velocity to correct for Doppler shift \cite{Bowen2020d}, an average parallel left-handed cyclotron spectrum $I(k_\parallel)$ is established. 

For each observed $\fpv$, a value of $\mathcal{H}$ is obtained numerically through Eq \ref{eq:H} using bi-Maxwellian and Hermite representations of $f(v_\perp, v_\parallel)$. For the distribution shown in Fig 2, a heating rate of $ 10^{-14}$ W/m$^3$ is found using the Hermite representation and $4 \times 10^{-15}$ W/m$^3$ using the drifting bi-Maxwellian spectrum. The measured $\mathcal{H}$ is similar to estimates of bulk ion heating due to turbulent dissipation at the spacecraft's location (30$R_\odot$) \cite{Chandran2019,Hellinger2013,Raja2021}.

Fig 3(a-d) shows the differential volumetric heating rate $\mathcal{H}$ as a function of resonant parallel velocity measured in each distribution function. The top panels, Fig 3(a-b), show positive $\mathcal{H}$, the ``heating'' rate, as a function of $v_\parallel$ and time for the bi-Maxwellian (a) and Hermite (b) representations. The bottom panels Fig 3(c-d) shows negative $\mathcal{H}$, the ``cooling'' rate over the interval as a function or resonant $v_\parallel$ due to emission of waves through instability. Fig 3(e) shows the net integrated $\mathcal{H}$ for each measured distribution function. The integrated $\mathcal{H}$ is uniformly positive, indicating that cyclotron waves present in the plasma are likely absorbed. There is very little cyclotron resonant emission from this plasma. However, the Hermite representation shows that cylotron-instability, when present, is focused at the parallel thermal speed,$v_{\parallel{th}}$.  The median heating rate is $3 \times10^{-15}$ W/m$^3$ for the bi-Maxwellian fits, and $1 \times10^{-14}$ W/m$^3$ for the Hermite representation. Using third order moments of the inertial range turbulence, \cite{Politano1998a}, a cascade rate of  4.7 $\times 10^{-14}$ W/m$^3$ is measured, which is consistent with previous measurements of the energy cascade rate at a similar radius \cite{Bandyopadhyay2020,Andres2022}. The measured cyclotron heating $\mathcal{H}$ ranges from approximately 10-20\% percent of the measured cascade rate. While uncertainties on the cascade rate exist due to the assumption of stationarity, isotropy, and homogeneity \cite{Politano1998a}, previous work has suggested that the cascade rate estimates through third-order moments are accurate to a factor of approximately two or three \cite{Stawarz2009,Osman2011PRL,Bandyopadhyay2020}. %Ensemble averaging of turbulent intervals observed by PSP suggests that at most an order of magnitude variation may occur \cite{Andres2022}.

 Observations from SPANi are partial measurements of $\fpv$ and are subject to both uncertainties and ongoing calibration work. However, during this interval, $\fpv$ is well resolved, and while uncertainties in $\fpv$ will dominate uncertainties in our measured heating rates, we argue that the measurements reliably suggest net energy transfer from waves to the particles. Specifically, the biMaxwellian fit removes fine-structure in $\fpv$ that is present in the measurements and replicated by the Hermite function; importantly, we find that removing fine structure (i.e. the bi-Maxwellian fit) produces heating rates of the same sign and order of magnitude as when fine structure in $\fpv$ is included. In essence, while fine structure in $\fpv$ is observed by SPAN, it neither drastically affects the order of magnitude, nor, importantly, the sign of the cyclotron heating rate, $\mathcal{H}$. However, these results highlight the importance of modeling fine structure and local gradients in $\fpv$, e.g. with Hermite functions, which may yield heating rates more than double those found by smoother approximations to $\fpv$, e.g. a bi-Maxwellian fit.

\begin{figure}[!]
    \centering
    \includegraphics[width=3.5in]{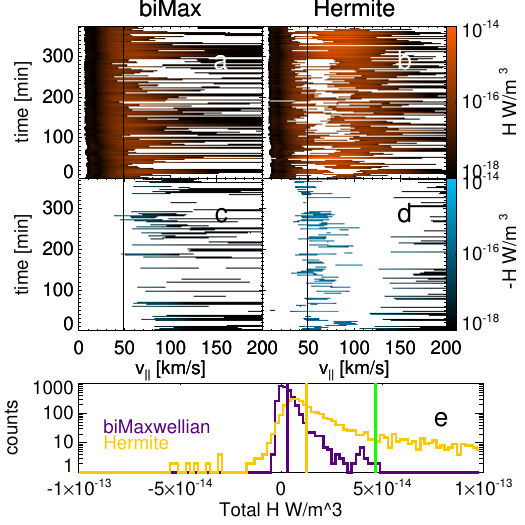}
   \caption{(a) Differential volumetric heating rates (positive $\mathcal{H}$) as a function of resonant parallel velocity computed from observed cyclotron spectra and drifting biMaxwellian fit to ion-distribution functions. The total heating rate is the sum over $v_\parallel$. Solid lines show the average $v_{\parallel{th}}$ for the interval (49 km/s). (b) Measured heating rates for a Hermite function approximation. (c-d) Corresponding cooling rates (negative $\mathcal{H}$). (e) Distribution of total $\mathcal{H}$ for biMaxwellian and Hermite representations; respective lines show median value of $\mathcal{H}$ The turbulent cascade rate estimated from third-order moments is shown in green.}
    \label{fig:3}
\end{figure}
      
\paragraph{Discussion}
Cyclotron resonance may play a significant role in shaping observed magnetic field spectra and distribution functions \cite{Goldstein1994,Leamon1998b,Leamon2000,Woodham2018,Telloni2019,MarschTu2001,He2015} observed in the solar wind. %Recent work has shown that QL diffusion likely broadens the proton beam through resonance with outward propagating right-handed modes \cite{KennelEngelmann1966,Verniero2021}.  Here, we show directly that application of QL diffusion to observed wave spectra and ion-distributions functions suggests the absorption of ion-cyclotron waves may lead to the bulk heating of the solar wind \cite{KennelEngelmann1966,IsenbergLee1996,Isenberg2001}. %Flattening along QL diffusion contours implies an equilibrium with respect to the Alfv\'{e}n/ion cyclotron instability; in our observations, flattening is only observed in the interpolated and drifting (two component) bi-Maxwellian approximations to the data, indicating that correctly understanding growth/damping rates through quasilinaer contours requires accurate descriptions of $\fpv$ that may not be possible with a single particle population.
While flattening along QL diffusion contours has been previously reported \cite{MarschTu2001,He2015, Verniero2021} our observations directly couple an observed spectrum of cyclotron waves to heating rates in measured distribution functions. Our measured heating rates are on the order of the measured energy cascade rates $( \sim 10^{-14}$ W/m$^3)$, obtained through third order moments of the observed turbulence  \cite{Politano1998a}; while this estimate may have significant uncertainty, we find good agreement with previous estimates of the cascade rate \cite{Bandyopadhyay2020,Andres2022}. We furthermore show that incorporating fine, nonthermal, structures in the distribution function using a Hermite functional decomposition introduces relatively little effect on the extent or sign of measured cyclotron heating. We thus argue that the measured levels of cyclotron heating provide significant evidence for the mediation of turbulent dissipation through cyclotron resonance that is potentially sufficient to power the solar wind \cite{Hellinger2013,Chandran2019,Raja2021}. Studying the radial scaling of the turbulent energy cascade alongside quasilinear cyclotron heating rates promises to further constrain cyclotron resonance as a dissipation process.

There are several sources of uncertainty in this analysis, rising predominantly in the estimate of the cascade rate and in the gradients of $\fpv$ due to limited resolution. Furthermore, if the occurrence rates of cyclotron waves decreases with $\theta_{vB}$, then the heating rate may be limited at oblique $\theta_{vB}$. Additionally, there is the potential that heating by oblique kinetic Alfv\'{e}n waves or oblique cyclotron waves may generate parallel cyclotron waves as a secondary process \cite{Chandran2010,Squire2022}. The spectrum of oblique cyclotron waves is difficult to distinguish due to the strong anisotropy of the background turbulence \cite{Bowen2020a}, though future work will explore signatures of oblique cyclotron resonance.  In any case, observed ion distributions are often flat along the quasi-parallel cyclotron diffusion contours and are rarely unstable to the growth of the waves. This flattening suggests that even if other physical processes contribute to bulk heating, the parallel-cyclotron resonance \cite{KennelEngelmann1966,IsenbergLee1996} plays a significant role in shaping the distribution functions. 
%Though omitted from the paper, analysis reveals significant statistical flattening of the distribution function along the diffusion contours indicating the role of cyclotron resonance \cite{KennelEngelmann1966,IsenbergLee1996}. 
%nearly exactly to steep magnetic spectra at the cyclotron resonance scale attributable to turbulent dissipation and heating \cite{Bowen2020c} %Intuitively, if no cyclotron emission is expected, then in order for the distribution to follow cyclotron resonant diffusion contours, heating rates should correspond well with requisite perpendicular heating rates for bulk heating.

The measured heating rate indicates a near total lack of cyclotron emission through instabilities; thus, the origin of cyclotron waves remains an important unresolved point. Our observations show that 95\% of the time left handed signatures are present, the net heating rate is positive, suggesting absorption of waves \cite{Vech2021}. We note the studied interval does not have strong cyclotron wave storms \cite{Bowen2020a,Verniero2020}, though application of our method to a similar interval with more significant cyclotron wave events similarly suggests net heating. There are two main possible physical origins for these Alfv\'{e}n/ion cyclotron waves. First, it is possible they are excited by beam instabilities \cite{Verscharen2013}, though recent work has suggested that dominant instability associated with the strong beam is associated with right-handed modes \cite{Verniero2020,Klein2021}. Second, they may be generated by turbulence, though canonical theories of Alfv\'{e}nic nonlinearity preferentially transport energy to large $k_\perp$ but not large $k_\parallel$ \cite{Shebalin1983,GS95}, which is a hurdle to the turbulent generation of cyclotron resonant waves. However, recent work suggests that imbalance, i.e.~the dominance of the outward Alfv\'{e}n mode, may prevent energy from cascading to kinetic scales \cite{Meyrand2021}. Fully kinetic simulations in the presence of such a barrier \cite{Squire2022} show the generation of quasi-parallel cyclotron modes, similar to those observed in the solar wind, providing a novel method for generating cyclotron waves that is consistent with a variety of observations \cite{Stawarz2010,Bale2019,Bowen2020a,Chen2020,Duan2020,Duan2021,Mozer2020,Bowen2020c,McManus2020}. Further work is required to specifically investigate the origin of the observed cyclotron waves and their connection to turbulence, though our observation of localized instability at the proton-core thermal speed (Fig \ref{fig:3}) may be a clue regarding the origin of the waves and their connection to the net heating measured in this study.

%; we note that this interval does not have strong cyclotron wave storms that may be driven by instabilities \cite{Bowen2020a,Verniero2020}. While we have focused on the signature of parallel cyclotron waves, the damping of oblique waves may play a role in heating \cite{Hollweg2002,Isenberg2011}; however, measured wave populations are known to be strongly parallel propagating, and the the oblique diffusion contours are steeper than the parallel contours \cite{Chandran2010b}, suggesting the observed distributions are likewise subject to oblique damping.
%A%dditionally, the damping of oblique cyclotron waves has been shown to drive $\fpv$ unstable to parallel-ICW modes \cite{Chandran2010b}, that are commonly measured in the solar wind, which may subsquently cause the distribution to flatten along the parallel-ICW QL diffusion curves. %The generation of parallel modes through the helicity barrier

This work explicitly shows that the measured distribution functions in the solar wind contain evidence of cyclotron resonant heating at a level that may power the expanding solar wind. These results are significant step towards understanding the underlying physics of collisionless heating and a kinetic description of astrophysical plasmas.

\begin{acknowledgments}
TAB  was supported by NASA PSP-GI Grant 80NSSC21K1771. KGK was supported by NASA ECIP Grant 80NSSC19K0912. BDGC was supported in part by NASA grant 80NSSC19K0829. JLV was supported by NASA grant NNH20ZDA001N-PSPGI. LDW was supported by STFC consolidated grant ST/S000364/1. DD acknowledges NSFC grant 42204166. The authors additionally acknowledge PSP/SWEAP \& FIELDS contract NNN06AA01C. 
\end{acknowledgments}

\providecommand{\noopsort}[1]{}\providecommand{\singleletter}[1]{#1}%\providecommand{\noopsort}[1]{}\providecommand{\singleletter}[1]{#1}%


\begin{thebibliography}{98}
\providecommand{\natexlab}[1]{#1}
\providecommand{\url}[1]{\texttt{#1}}
\expandafter\ifx\csname urlstyle\endcsname\relax
  \providecommand{\doi}[1]{doi: #1}\else
  \providecommand{\doi}{doi: \begingroup \urlstyle{rm}\Url}\fi

\bibitem[{Marsch}(2006)]{Marsch2006}
Eckart {Marsch}.
\newblock {Kinetic Physics of the Solar Corona and Solar Wind}.
\newblock \emph{Living Reviews in Solar Physics}, 3\penalty0 (1):\penalty0 1,
  July 2006.
\newblock \doi{10.12942/lrsp-2006-1}.

\bibitem[{Coleman}(1968)]{Coleman1968}
Jr. {Coleman}, Paul~J.
\newblock {Turbulence, Viscosity, and Dissipation in the Solar-Wind Plasma}.
\newblock \emph{The Astrophysical Journal}, 153:\penalty0 371, Aug 1968.
\newblock \doi{10.1086/149674}.

\bibitem[{Matthaeus} and {Goldstein}(1982)]{MatthaeusGoldstein1982}
W.~H. {Matthaeus} and M.~L. {Goldstein}.
\newblock {Measurement of the rugged invariants of magnetohydrodynamic
  turbulence in the solar wind}.
\newblock \emph{Journal of Geophysical Research}, 87:\penalty0 6011--6028, Aug
  1982.
\newblock \doi{10.1029/JA087iA08p06011}.

\bibitem[{Cranmer} and {van Ballegooijen}(2003)]{CranmervanBallegooijen2003}
S.~R. {Cranmer} and A.~A. {van Ballegooijen}.
\newblock {Alfv{\'e}nic Turbulence in the Extended Solar Corona: Kinetic
  Effects and Proton Heating}.
\newblock \emph{The Astrophysical Journal}, 594\penalty0 (1):\penalty0
  573--591, Sep 2003.
\newblock \doi{10.1086/376777}.

\bibitem[{Chandran} et~al.(2010{\natexlab{a}}){Chandran}, {Li}, {Rogers},
  {Quataert}, and {Germaschewski}]{Chandran2010a}
Benjamin D.~G. {Chandran}, Bo~{Li}, Barrett~N. {Rogers}, Eliot {Quataert}, and
  Kai {Germaschewski}.
\newblock {Perpendicular Ion Heating by Low-frequency Alfv{\'e}n-wave
  Turbulence in the Solar Wind}.
\newblock \emph{The Astrophysical Journal}, 720\penalty0 (1):\penalty0
  503--515, Sep 2010{\natexlab{a}}.
\newblock \doi{10.1088/0004-637X/720/1/503}.

\bibitem[{Barnes}(1966)]{Barnes1966}
A.~{Barnes}.
\newblock {Collisionless Damping of Hydromagnetic Waves}.
\newblock \emph{Physics of Fluids}, 9:\penalty0 1483--1495, August 1966.
\newblock \doi{10.1063/1.1761882}.

\bibitem[{Barnes}(1979)]{Barnes1979}
A.~{Barnes}.
\newblock \emph{{Hydromagnetic waves and turbulence in the solar wind}},
  volume~1, pages 249--319.
\newblock 1979.

\bibitem[{Goldstein} et~al.(1994){Goldstein}, {Roberts}, and
  {Fitch}]{Goldstein1994}
M.~L. {Goldstein}, D.~A. {Roberts}, and C.~A. {Fitch}.
\newblock {Properties of the fluctuating magnetic helicity in the inertial and
  dissipation ranges of solar wind turbulence}.
\newblock \emph{Journal of Geophysical Research}, 99\penalty0 (A6):\penalty0
  11519--11538, Jun 1994.
\newblock \doi{10.1029/94JA00789}.

\bibitem[{Leamon} et~al.(1998{\natexlab{a}}){Leamon}, {Matthaeus}, {Smith}, and
  {Wong}]{Leamon1998b}
Robert~J. {Leamon}, William~H. {Matthaeus}, Charles~W. {Smith}, and Hung~K.
  {Wong}.
\newblock {Contribution of Cyclotron-resonant Damping to Kinetic Dissipation of
  Interplanetary Turbulence}.
\newblock \emph{The Astrophysical Journal Letters}, 507\penalty0 (2):\penalty0
  L181--L184, Nov 1998{\natexlab{a}}.
\newblock \doi{10.1086/311698}.

\bibitem[{Howes} et~al.(2011){Howes}, {Tenbarge}, {Dorland}, {Quataert},
  {Schekochihin}, {Numata}, and {Tatsuno}]{Howes2011}
G.~G. {Howes}, J.~M. {Tenbarge}, W.~{Dorland}, E.~{Quataert}, A.~A.
  {Schekochihin}, R.~{Numata}, and T.~{Tatsuno}.
\newblock {Gyrokinetic Simulations of Solar Wind Turbulence from Ion to
  Electron Scales}.
\newblock \emph{Physical Review Letters}, 107\penalty0 (3):\penalty0 035004,
  Jul 2011.
\newblock \doi{10.1103/PhysRevLett.107.035004}.

\bibitem[{Chandran} et~al.(2013){Chandran}, {Verscharen}, {Quataert}, {Kasper},
  {Isenberg}, and {Bourouaine}]{Chandran2013}
B.~D.~G. {Chandran}, D.~{Verscharen}, E.~{Quataert}, J.~C. {Kasper}, P.~A.
  {Isenberg}, and S.~{Bourouaine}.
\newblock {Stochastic Heating, Differential Flow, and the Alpha-to-proton
  Temperature Ratio in the Solar Wind}.
\newblock \emph{ApJ}, 776\penalty0 (1):\penalty0 45, October 2013.
\newblock \doi{10.1088/0004-637X/776/1/45}.

\bibitem[{Mallet} et~al.(2017){Mallet}, {Schekochihin}, and {Chand
  ran}]{Mallet2017}
Alfred {Mallet}, Alexander~A. {Schekochihin}, and Benjamin D.~G. {Chand ran}.
\newblock {Disruption of Alfv{\'e}nic turbulence by magnetic reconnection in a
  collisionless plasma}.
\newblock \emph{Journal of Plasma Physics}, 83\penalty0 (6):\penalty0
  905830609, Dec 2017.
\newblock \doi{10.1017/S0022377817000812}.

\bibitem[{Loureiro} and {Boldyrev}(2017)]{Loureiro2017a}
Nuno~F. {Loureiro} and Stanislav {Boldyrev}.
\newblock {Role of Magnetic Reconnection in Magnetohydrodynamic Turbulence}.
\newblock \emph{Physical Review Letters}, 118\penalty0 (24):\penalty0 245101,
  Jun 2017.
\newblock \doi{10.1103/PhysRevLett.118.245101}.

\bibitem[{Quataert} and {Gruzinov}(1999)]{QuataertGruzinov1999}
Eliot {Quataert} and Andrei {Gruzinov}.
\newblock {Turbulence and Particle Heating in Advection-dominated Accretion
  Flows}.
\newblock \emph{The Astrophysical Journal}, 520\penalty0 (1):\penalty0
  248--255, Jul 1999.
\newblock \doi{10.1086/307423}.

\bibitem[{Chandran} et~al.(2010{\natexlab{b}}){Chandran}, {Pongkitiwanichakul},
  {Isenberg}, {Lee}, {Markovskii}, {Hollweg}, and {Vasquez}]{Chandran2010b}
Benjamin D.~G. {Chandran}, Peera {Pongkitiwanichakul}, Philip~A. {Isenberg},
  Martin~A. {Lee}, Sergei~A. {Markovskii}, Joseph~V. {Hollweg}, and Bernard~J.
  {Vasquez}.
\newblock {Resonant Interactions Between Protons and Oblique
  Alfv{\'e}n/Ion-cyclotron Waves in the Solar Corona and Solar Flares}.
\newblock \emph{ApJ}, 722\penalty0 (1):\penalty0 710--720, October
  2010{\natexlab{b}}.
\newblock \doi{10.1088/0004-637X/722/1/710}.

\bibitem[{Chandran} et~al.(2011){Chandran}, {Dennis}, {Quataert}, and
  {Bale}]{Chandran2011}
Benjamin D.~G. {Chandran}, Timothy~J. {Dennis}, Eliot {Quataert}, and Stuart~D.
  {Bale}.
\newblock {Incorporating Kinetic Physics into a Two-fluid Solar-wind Model with
  Temperature Anisotropy and Low-frequency Alfv{\'e}n-wave Turbulence}.
\newblock \emph{The Astrophysical Journal}, 743\penalty0 (2):\penalty0 197, Dec
  2011.
\newblock \doi{10.1088/0004-637X/743/2/197}.

\bibitem[{Alexandrova} et~al.(2012){Alexandrova}, {Lacombe}, {Mangeney},
  {Grappin}, and {Maksimovic}]{Alexandrova2012}
O.~{Alexandrova}, C.~{Lacombe}, A.~{Mangeney}, R.~{Grappin}, and
  M.~{Maksimovic}.
\newblock {Solar Wind Turbulent Spectrum at Plasma Kinetic Scales}.
\newblock \emph{ApJ}, 760\penalty0 (2):\penalty0 121, December 2012.
\newblock \doi{10.1088/0004-637X/760/2/121}.

\bibitem[{Richardson} et~al.(1995){Richardson}, {Paularena}, {Lazarus}, and
  {Belcher}]{Richardson1995}
John~D. {Richardson}, Karolen~I. {Paularena}, Alan~J. {Lazarus}, and John~W.
  {Belcher}.
\newblock {Radial evolution of the solar wind from IMP 8 to Voyager 2}.
\newblock \emph{Geophys. Research Letters}, 22\penalty0 (4):\penalty0 325--328,
  Feb 1995.
\newblock \doi{10.1029/94GL03273}.

\bibitem[{Matteini} et~al.(2007){Matteini}, {Landi}, {Hellinger}, {Pantellini},
  {Maksimovic}, {Velli}, {Goldstein}, and {Marsch}]{Matteini2007}
Lorenzo {Matteini}, Simone {Landi}, Petr {Hellinger}, Filippo {Pantellini},
  Milan {Maksimovic}, Marco {Velli}, Bruce~E. {Goldstein}, and Eckart {Marsch}.
\newblock {Evolution of the solar wind proton temperature anisotropy from 0.3
  to 2.5 AU}.
\newblock \emph{GRL}, 34\penalty0 (20):\penalty0 L20105, October 2007.
\newblock \doi{10.1029/2007GL030920}.

\bibitem[{Quataert}(1998)]{Quataert1998}
Eliot {Quataert}.
\newblock {Particle Heating by Alfv{\'e}nic Turbulence in Hot Accretion Flows}.
\newblock \emph{The Astrophysical Journal}, 500\penalty0 (2):\penalty0
  978--991, Jun 1998.
\newblock \doi{10.1086/305770}.

\bibitem[{Leamon} et~al.(1998{\natexlab{b}}){Leamon}, {Smith}, {Ness},
  {Matthaeus}, and {Wong}]{Leamon1998a}
Robert~J. {Leamon}, Charles~W. {Smith}, Norman~F. {Ness}, William~H.
  {Matthaeus}, and Hung~K. {Wong}.
\newblock {Observational constraints on the dynamics of the interplanetary
  magnetic field dissipation range}.
\newblock \emph{Journal of Geophysical Research}, 103\penalty0 (A3):\penalty0
  4775--4788, Mar 1998{\natexlab{b}}.
\newblock \doi{10.1029/97JA03394}.

\bibitem[{Gary}(1999)]{Gary1999}
S.~Peter {Gary}.
\newblock {Collisionless dissipation wavenumber: Linear theory}.
\newblock \emph{JGR}, 104\penalty0 (A4):\penalty0 6759--6762, April 1999.
\newblock \doi{10.1029/1998JA900161}.

\bibitem[{Leamon} et~al.(2000){Leamon}, {Matthaeus}, {Smith}, {Zank}, {Mullan},
  and {Oughton}]{Leamon2000}
R.~J. {Leamon}, W.~H. {Matthaeus}, C.~W. {Smith}, G.~P. {Zank}, D.~J. {Mullan},
  and S.~{Oughton}.
\newblock {MHD-driven Kinetic Dissipation in the Solar Wind and Corona}.
\newblock \emph{The Astrophysical Journal}, 537\penalty0 (2):\penalty0
  1054--1062, Jul 2000.
\newblock \doi{10.1086/309059}.

\bibitem[{Howes} et~al.(2008){Howes}, {Cowley}, {Dorland}, {Hammett},
  {Quataert}, and {Schekochihin}]{Howes2008}
G.~G. {Howes}, S.~C. {Cowley}, W.~{Dorland}, G.~W. {Hammett}, E.~{Quataert},
  and A.~A. {Schekochihin}.
\newblock {A model of turbulence in magnetized plasmas: Implications for the
  dissipation range in the solar wind}.
\newblock \emph{Journal of Geophysical Research (Space Physics)}, 113\penalty0
  (A5):\penalty0 A05103, May 2008.
\newblock \doi{10.1029/2007JA012665}.

\bibitem[{Stix}(1992)]{Stix1992}
Thomas~Howard {Stix}.
\newblock \emph{{Waves in plasmas}}.
\newblock 1992.

\bibitem[{Hollweg} and {Johnson}(1988)]{HollwegJohnson1988}
Joseph~V. {Hollweg} and Walter {Johnson}.
\newblock {Transition region, corona, and solar wind in coronal holes: Some
  two-fluid models}.
\newblock \emph{JGR}, 93\penalty0 (A9):\penalty0 9547--9554, September 1988.
\newblock \doi{10.1029/JA093iA09p09547}.

\bibitem[{Tu} and {Marsch}(1997)]{TuMarsch1997}
C.~Y. {Tu} and E.~{Marsch}.
\newblock {Two-Fluid Model for Heating of the Solar Corona and Acceleration of
  the Solar Wind by High-Frequency Alfven Waves}.
\newblock \emph{Sol Phys}, 171\penalty0 (2):\penalty0 363--391, April 1997.
\newblock \doi{10.1023/A:1004968327196}.

\bibitem[{Cranmer}(2000)]{Cranmer2000}
Steven~R. {Cranmer}.
\newblock {Ion Cyclotron Wave Dissipation in the Solar Corona: The Summed
  Effect of More than 2000 Ion Species}.
\newblock \emph{The Astrophysical Journal}, 532\penalty0 (2):\penalty0
  1197--1208, Apr 2000.
\newblock \doi{10.1086/308620}.

\bibitem[{Hollweg} and {Isenberg}(2002)]{HollwegIsenberg2002}
Joseph~V. {Hollweg} and Philip~A. {Isenberg}.
\newblock {Generation of the fast solar wind: A review with emphasis on the
  resonant cyclotron interaction}.
\newblock \emph{Journal of Geophysical Research (Space Physics)}, 107\penalty0
  (A7):\penalty0 1147, July 2002.
\newblock \doi{10.1029/2001JA000270}.

\bibitem[{Cranmer}(2014)]{Cranmer2014}
Steven~R. {Cranmer}.
\newblock {Ensemble Simulations of Proton Heating in the Solar Wind via
  Turbulence and Ion Cyclotron Resonance}.
\newblock \emph{ApJS}, 213\penalty0 (1):\penalty0 16, July 2014.
\newblock \doi{10.1088/0067-0049/213/1/16}.

\bibitem[{Kohl} et~al.(1997){Kohl}, {Noci}, {Antonucci}, {Tondello}, {Huber},
  {Gardner}, {Nicolosi}, {Strachan}, {Fineschi}, {Raymond}, {Romoli},
  {Spadaro}, {Panasyuk}, {Siegmund}, {Benna}, {Ciaravella}, {Cranmer},
  {Giordano}, {Karovska}, {Martin}, {Michels}, {Modigliani}, {Naletto},
  {Pernechele}, {Poletto}, and {Smith}]{Kohl1997}
J.~L. {Kohl}, G.~{Noci}, E.~{Antonucci}, G.~{Tondello}, M.~C.~E. {Huber}, L.~D.
  {Gardner}, P.~{Nicolosi}, L.~{Strachan}, S.~{Fineschi}, J.~C. {Raymond},
  M.~{Romoli}, D.~{Spadaro}, A.~{Panasyuk}, O.~H.~W. {Siegmund}, C.~{Benna},
  A.~{Ciaravella}, S.~R. {Cranmer}, S.~{Giordano}, M.~{Karovska}, R.~{Martin},
  J.~{Michels}, A.~{Modigliani}, G.~{Naletto}, C.~{Pernechele}, G.~{Poletto},
  and P.~L. {Smith}.
\newblock {First Results from the SOHO Ultraviolet Coronagraph Spectrometer}.
\newblock \emph{Sol Phys}, 175\penalty0 (2):\penalty0 613--644, October 1997.
\newblock \doi{10.1023/A:1004903206467}.

\bibitem[{Kohl} et~al.(1998){Kohl}, {Noci}, {Antonucci}, {Tondello}, {Huber},
  {Cranmer}, {Strachan}, {Panasyuk}, {Gardner}, {Romoli}, {Fineschi},
  {Dobrzycka}, {Raymond}, {Nicolosi}, {Siegmund}, {Spadaro}, {Benna},
  {Ciaravella}, {Giordano}, {Habbal}, {Karovska}, {Li}, {Martin}, {Michels},
  {Modigliani}, {Naletto}, {O'Neal}, {Pernechele}, {Poletto}, {Smith}, and
  {Suleiman}]{Kohl1998}
J.~L. {Kohl}, G.~{Noci}, E.~{Antonucci}, G.~{Tondello}, M.~C.~E. {Huber}, S.~R.
  {Cranmer}, L.~{Strachan}, A.~V. {Panasyuk}, L.~D. {Gardner}, M.~{Romoli},
  S.~{Fineschi}, D.~{Dobrzycka}, J.~C. {Raymond}, P.~{Nicolosi}, O.~H.~W.
  {Siegmund}, D.~{Spadaro}, C.~{Benna}, A.~{Ciaravella}, S.~{Giordano}, S.~R.
  {Habbal}, M.~{Karovska}, X.~{Li}, R.~{Martin}, J.~G. {Michels},
  A.~{Modigliani}, G.~{Naletto}, R.~H. {O'Neal}, C.~{Pernechele}, G.~{Poletto},
  P.~L. {Smith}, and R.~M. {Suleiman}.
\newblock {UVCS/SOHO Empirical Determinations of Anisotropic Velocity
  Distributions in the Solar Corona}.
\newblock \emph{ApJL}, 501\penalty0 (1):\penalty0 L127--L131, July 1998.
\newblock \doi{10.1086/311434}.

\bibitem[{Cranmer} et~al.(1999){Cranmer}, {Field}, and {Kohl}]{Cranmer1999}
Steven~R. {Cranmer}, George~B. {Field}, and John~L. {Kohl}.
\newblock {Spectroscopic Constraints on Models of Ion Cyclotron Resonance
  Heating in the Polar Solar Corona and High-Speed Solar Wind}.
\newblock \emph{ApJ}, 518\penalty0 (2):\penalty0 937--947, June 1999.
\newblock \doi{10.1086/307330}.

\bibitem[{Jian} et~al.(2009){Jian}, {Russell}, {Luhmann}, {Strangeway},
  {Leisner}, and {Galvin}]{Jian2009}
Lan~K. {Jian}, Christopher~T. {Russell}, Janet~G. {Luhmann}, Robert~J.
  {Strangeway}, Jared~S. {Leisner}, and Antoinette~B. {Galvin}.
\newblock {Ion Cyclotron Waves in the Solar Wind Observed by STEREO Near 1 AU}.
\newblock \emph{ApJL}, 701\penalty0 (2):\penalty0 L105--L109, August 2009.
\newblock \doi{10.1088/0004-637X/701/2/L105}.

\bibitem[{Podesta} and {Gary}(2011)]{Podesta2011}
J.~J. {Podesta} and S.~P. {Gary}.
\newblock {Magnetic Helicity Spectrum of Solar Wind Fluctuations as a Function
  of the Angle with Respect to the Local Mean Magnetic Field}.
\newblock \emph{The Astrophysical Journal}, 734\penalty0 (1):\penalty0 15, Jun
  2011.
\newblock \doi{10.1088/0004-637X/734/1/15}.

\bibitem[{He} et~al.(2011){He}, {Marsch}, {Tu}, {Yao}, and {Tian}]{He2011}
Jiansen {He}, Eckart {Marsch}, Chuanyi {Tu}, Shuo {Yao}, and Hui {Tian}.
\newblock {Possible Evidence of Alfv{\'e}n-cyclotron Waves in the Angle
  Distribution of Magnetic Helicity of Solar Wind Turbulence}.
\newblock \emph{The Astrophysical Journal}, 731\penalty0 (2):\penalty0 85, Apr
  2011.
\newblock \doi{10.1088/0004-637X/731/2/85}.

\bibitem[{Wicks} et~al.(2016){Wicks}, {Alexander}, {Stevens}, {Wilson}, {Moya},
  {Vi{\~n}as}, {Jian}, {Roberts}, {O'Modhrain}, {Gilbert}, and
  {Zurbuchen}]{Wicks2016}
R.~T. {Wicks}, R.~L. {Alexander}, M.~{Stevens}, III {Wilson}, L.~B., P.~S.
  {Moya}, A.~{Vi{\~n}as}, L.~K. {Jian}, D.~A. {Roberts}, S.~{O'Modhrain}, J.~A.
  {Gilbert}, and T.~H. {Zurbuchen}.
\newblock {A Proton-cyclotron Wave Storm Generated by Unstable Proton
  Distribution Functions in the Solar Wind}.
\newblock \emph{ApJ}, 819\penalty0 (1):\penalty0 6, March 2016.
\newblock \doi{10.3847/0004-637X/819/1/6}.

\bibitem[{Woodham} et~al.(2019){Woodham}, {Wicks}, {Verscharen}, {Owen},
  {Maruca}, and {Alterman}]{Woodham2019}
Lloyd~D. {Woodham}, Robert~T. {Wicks}, Daniel {Verscharen}, Christopher~J.
  {Owen}, Bennett~A. {Maruca}, and Benjamin~L. {Alterman}.
\newblock {Parallel-propagating Fluctuations at Proton-kinetic Scales in the
  Solar Wind Are Dominated By Kinetic Instabilities}.
\newblock \emph{ApJL}, 884\penalty0 (2):\penalty0 L53, October 2019.
\newblock \doi{10.3847/2041-8213/ab4adc}.

\bibitem[Bale et~al.(2019)Bale, Badman, Bonnell, Bowen, Burgess, Case, Cattell,
  Chandran, Chaston, Chen, et~al.]{Bale2019}
SD~Bale, ST~Badman, JW~Bonnell, TA~Bowen, D~Burgess, AW~Case, CA~Cattell, BDG
  Chandran, CC~Chaston, CHK Chen, et~al.
\newblock Highly structured slow solar wind emerging from an equatorial coronal
  hole.
\newblock \emph{Nature}, pages 1--6, 2019.

\bibitem[{Bowen} et~al.(2020{\natexlab{a}}){Bowen}, {Mallet}, {Huang}, {Klein},
  {Malaspina}, {Stevens}, {Bale}, {Bonnell}, {Case}, {Chandran}, {Chaston},
  {Chen}, {Dudok de Wit}, {Goetz}, {Harvey}, {Howes}, {Kasper}, {Korreck},
  {Larson}, {Livi}, {MacDowall}, {McManus}, {Pulupa}, {Verniero}, and
  {Whittlesey}]{Bowen2020a}
Trevor~A. {Bowen}, Alfred {Mallet}, Jia {Huang}, Kristopher~G. {Klein},
  David~M. {Malaspina}, Michael {Stevens}, Stuart~D. {Bale}, J.~W. {Bonnell},
  Anthony~W. {Case}, Benjamin D.~G. {Chandran}, C.~C. {Chaston}, Christopher
  H.~K. {Chen}, Thierry {Dudok de Wit}, Keith {Goetz}, Peter~R. {Harvey},
  Gregory~G. {Howes}, J.~C. {Kasper}, Kelly~E. {Korreck}, Davin {Larson},
  Roberto {Livi}, Robert~J. {MacDowall}, Michael~D. {McManus}, Marc {Pulupa},
  J.~L. {Verniero}, and Phyllis {Whittlesey}.
\newblock {Ion-scale Electromagnetic Waves in the Inner Heliosphere}.
\newblock \emph{ApJS}, 246\penalty0 (2):\penalty0 66, February
  2020{\natexlab{a}}.
\newblock \doi{10.3847/1538-4365/ab6c65}.

\bibitem[{Isenberg}(1990)]{Isenberg1990}
Philip~A. {Isenberg}.
\newblock {Investigations of a turbulence-driven solar wind model}.
\newblock \emph{JGR}, 95\penalty0 (A5):\penalty0 6437--6442, May 1990.
\newblock \doi{10.1029/JA095iA05p06437}.

\bibitem[{Woodham} et~al.(2018){Woodham}, {Wicks}, {Verscharen}, and
  {Owen}]{Woodham2018}
Lloyd~D. {Woodham}, Robert~T. {Wicks}, Daniel {Verscharen}, and Christopher~J.
  {Owen}.
\newblock {The Role of Proton Cyclotron Resonance as a Dissipation Mechanism in
  Solar Wind Turbulence: A Statistical Study at Ion-kinetic Scales}.
\newblock \emph{The Astrophysical Journal}, 856\penalty0 (1):\penalty0 49, Mar
  2018.
\newblock \doi{10.3847/1538-4357/aab03d}.

\bibitem[{Kennel} and {Engelmann}(1966)]{KennelEngelmann1966}
C.~F. {Kennel} and F.~{Engelmann}.
\newblock {Velocity Space Diffusion from Weak Plasma Turbulence in a Magnetic
  Field}.
\newblock \emph{Physics of Fluids}, 9\penalty0 (12):\penalty0 2377--2388,
  December 1966.
\newblock \doi{10.1063/1.1761629}.

\bibitem[{Isenberg} and {Lee}(1996)]{IsenbergLee1996}
Philip~A. {Isenberg} and Martin~A. {Lee}.
\newblock {A dispersive analysis of bispherical pickup ion distributions}.
\newblock \emph{JGR}, 101\penalty0 (A5):\penalty0 11055--11066, May 1996.
\newblock \doi{10.1029/96JA00293}.

\bibitem[{Isenberg}(2001)]{Isenberg2001}
Philip~A. {Isenberg}.
\newblock {The kinetic shell model of coronal heating and acceleration by ion
  cyclotron waves: 2. Inward and outward propagating waves}.
\newblock \emph{JGR}, 106\penalty0 (A12):\penalty0 29249--29260, December 2001.
\newblock \doi{10.1029/2001JA000176}.

\bibitem[{Marsch} and {Tu}(2001)]{MarschTu2001}
E.~{Marsch} and C.~Y. {Tu}.
\newblock {Evidence for pitch angle diffusion of solar wind protons in
  resonance with cyclotron waves}.
\newblock \emph{JGR}, 106\penalty0 (A5):\penalty0 8357--8362, May 2001.
\newblock \doi{10.1029/2000JA000414}.

\bibitem[{Marsch} and {Bourouaine}(2011)]{MarschBourouaine2011}
E.~{Marsch} and S.~{Bourouaine}.
\newblock {Velocity-space diffusion of solar wind protons in oblique waves and
  weak turbulence}.
\newblock \emph{Annales Geophysicae}, 29\penalty0 (11):\penalty0 2089--2099,
  November 2011.
\newblock \doi{10.5194/angeo-29-2089-2011}.

\bibitem[{He} et~al.(2015){He}, {Wang}, {Tu}, {Marsch}, and {Zong}]{He2015}
Jiansen {He}, Linghua {Wang}, Chuanyi {Tu}, Eckart {Marsch}, and Qiugang
  {Zong}.
\newblock {Evidence of Landau and Cyclotron Resonance between Protons and
  Kinetic Waves in Solar Wind Turbulence}.
\newblock \emph{ApJ}, 800\penalty0 (2):\penalty0 L31, February 2015.
\newblock \doi{10.1088/2041-8205/800/2/L31}.

\bibitem[{Gary} et~al.(2001){Gary}, {Skoug}, {Steinberg}, and
  {Smith}]{Gary2001}
S.~Peter {Gary}, Ruth~M. {Skoug}, John~T. {Steinberg}, and Charles~W. {Smith}.
\newblock {Proton temperature anisotropy constraint in the solar wind: ACE
  observations}.
\newblock \emph{GRL}, 28\penalty0 (14):\penalty0 2759--2762, January 2001.
\newblock \doi{10.1029/2001GL013165}.

\bibitem[{Gary} et~al.(2016){Gary}, {Jian}, {Broiles}, {Stevens}, {Podesta},
  and {Kasper}]{Gary2016}
S.~Peter {Gary}, Lan~K. {Jian}, Thomas~W. {Broiles}, Michael~L. {Stevens},
  John~J. {Podesta}, and Justin~C. {Kasper}.
\newblock {Ion-driven instabilities in the solar wind: Wind observations of 19
  March 2005}.
\newblock \emph{Journal of Geophysical Research (Space Physics)}, 121\penalty0
  (1):\penalty0 30--41, January 2016.
\newblock \doi{10.1002/2015JA021935}.

\bibitem[{Kasper} et~al.(2008){Kasper}, {Lazarus}, and {Gary}]{Kasper2008}
J.~C. {Kasper}, A.~J. {Lazarus}, and S.~P. {Gary}.
\newblock {Hot Solar-Wind Helium: Direct Evidence for Local Heating by
  Alfv{\'e}n-Cyclotron Dissipation}.
\newblock \emph{PRL}, 101\penalty0 (26):\penalty0 261103, December 2008.
\newblock \doi{10.1103/PhysRevLett.101.261103}.

\bibitem[{Telloni} et~al.(2019){Telloni}, {Carbone}, {Bruno}, {Zank},
  {Sorriso-Valvo}, and {Mancuso}]{Telloni2019}
Daniele {Telloni}, Francesco {Carbone}, Roberto {Bruno}, Gary~P. {Zank}, Luca
  {Sorriso-Valvo}, and Salvatore {Mancuso}.
\newblock {Ion Cyclotron Waves in Field-aligned Solar Wind Turbulence}.
\newblock \emph{ApJL}, 885\penalty0 (1):\penalty0 L5, November 2019.
\newblock \doi{10.3847/2041-8213/ab4c44}.

\bibitem[{Vech} et~al.(2021){Vech}, {Martinovi{\'c}}, {Klein}, {Malaspina},
  {Bowen}, {Verniero}, {Paulson}, {Dudok de Wit}, {Kasper}, {Huang}, {Stevens},
  {Case}, {Korreck}, {Mozer}, {Goodrich}, {Bale}, {Whittlesey}, {Livi},
  {Larson}, {Pulupa}, {Bonnell}, {Harvey}, {Goetz}, and {MacDowall}]{Vech2021}
D.~{Vech}, M.~M. {Martinovi{\'c}}, K.~G. {Klein}, D.~M. {Malaspina}, T.~A.
  {Bowen}, J.~L. {Verniero}, K.~{Paulson}, T.~{Dudok de Wit}, J.~C. {Kasper},
  J.~{Huang}, M.~L. {Stevens}, A.~W. {Case}, K.~{Korreck}, F.~S. {Mozer}, K.~A.
  {Goodrich}, S.~D. {Bale}, P.~L. {Whittlesey}, R.~{Livi}, D.~E. {Larson},
  M.~{Pulupa}, J.~{Bonnell}, P.~{Harvey}, K.~{Goetz}, and R.~{MacDowall}.
\newblock {Wave-particle energy transfer directly observed in an ion cyclotron
  wave}.
\newblock \emph{A\&AP}, 650:\penalty0 A10, June 2021.
\newblock \doi{10.1051/0004-6361/202039296}.

\bibitem[Fox et~al.(2016)Fox, Velli, Bale, Decker, Driesman, Howard, Kasper,
  Kinnison, Kusterer, Lario, Lockwood, McComas, Raouafi, and Szabo]{Fox2016}
N.~J. Fox, M.~C. Velli, S.~D. Bale, R.~Decker, A.~Driesman, R.~A. Howard, J.~C.
  Kasper, J.~Kinnison, M.~Kusterer, D.~Lario, M.~K. Lockwood, D.~J. McComas,
  N.~E. Raouafi, and A.~Szabo.
\newblock The solar probe plus mission: Humanity's first visit to our star.
\newblock \emph{Space Science Reviews}, 204\penalty0 (1):\penalty0 7--48, Dec
  2016.
\newblock ISSN 1572-9672.
\newblock \doi{10.1007/s11214-015-0211-6}.
\newblock URL \url{https://doi.org/10.1007/s11214-015-0211-6}.

\bibitem[{Bale} et~al.(2016){Bale}, {Goetz}, {Harvey}, {Turin}, {Bonnell},
  {Dudok de Wit}, {Ergun}, {MacDowall}, {Pulupa}, {Andre}, {Bolton},
  {Bougeret}, {Bowen}, {Burgess}, {Cattell}, {Chandran}, {Chaston}, {Chen},
  {Choi}, {Connerney}, {Cranmer}, {Diaz-Aguado}, {Donakowski}, {Drake},
  {Farrell}, {Fergeau}, {Fermin}, {Fischer}, {Fox}, {Glaser}, {Goldstein},
  {Gordon}, {Hanson}, {Harris}, {Hayes}, {Hinze}, {Hollweg}, {Horbury},
  {Howard}, {Hoxie}, {Jannet}, {Karlsson}, {Kasper}, {Kellogg}, {Kien},
  {Klimchuk}, {Krasnoselskikh}, {Krucker}, {Lynch}, {Maksimovic}, {Malaspina},
  {Marker}, {Martin}, {Martinez-Oliveros}, {McCauley}, {McComas}, {McDonald},
  {Meyer-Vernet}, {Moncuquet}, {Monson}, {Mozer}, {Murphy}, {Odom},
  {Oliverson}, {Olson}, {Parker}, {Pankow}, {Phan}, {Quataert}, {Quinn},
  {Ruplin}, {Salem}, {Seitz}, {Sheppard}, {Siy}, {Stevens}, {Summers}, {Szabo},
  {Timofeeva}, {Vaivads}, {Velli}, {Yehle}, {Werthimer}, and
  {Wygant}]{Bale2016}
S.~D. {Bale}, K.~{Goetz}, P.~R. {Harvey}, P.~{Turin}, J.~W. {Bonnell},
  T.~{Dudok de Wit}, R.~E. {Ergun}, R.~J. {MacDowall}, M.~{Pulupa}, M.~{Andre},
  M.~{Bolton}, J.-L. {Bougeret}, T.~A. {Bowen}, D.~{Burgess}, C.~A. {Cattell},
  B.~D.~G. {Chandran}, C.~C. {Chaston}, C.~H.~K. {Chen}, M.~K. {Choi}, J.~E.
  {Connerney}, S.~{Cranmer}, M.~{Diaz-Aguado}, W.~{Donakowski}, J.~F. {Drake},
  W.~M. {Farrell}, P.~{Fergeau}, J.~{Fermin}, J.~{Fischer}, N.~{Fox},
  D.~{Glaser}, M.~{Goldstein}, D.~{Gordon}, E.~{Hanson}, S.~E. {Harris}, L.~M.
  {Hayes}, J.~J. {Hinze}, J.~V. {Hollweg}, T.~S. {Horbury}, R.~A. {Howard},
  V.~{Hoxie}, G.~{Jannet}, M.~{Karlsson}, J.~C. {Kasper}, P.~J. {Kellogg},
  M.~{Kien}, J.~A. {Klimchuk}, V.~V. {Krasnoselskikh}, S.~{Krucker}, J.~J.
  {Lynch}, M.~{Maksimovic}, D.~M. {Malaspina}, S.~{Marker}, P.~{Martin},
  J.~{Martinez-Oliveros}, J.~{McCauley}, D.~J. {McComas}, T.~{McDonald},
  N.~{Meyer-Vernet}, M.~{Moncuquet}, S.~J. {Monson}, F.~S. {Mozer}, S.~D.
  {Murphy}, J.~{Odom}, R.~{Oliverson}, J.~{Olson}, E.~N. {Parker}, D.~{Pankow},
  T.~{Phan}, E.~{Quataert}, T.~{Quinn}, S.~W. {Ruplin}, C.~{Salem}, D.~{Seitz},
  D.~A. {Sheppard}, A.~{Siy}, K.~{Stevens}, D.~{Summers}, A.~{Szabo},
  M.~{Timofeeva}, A.~{Vaivads}, M.~{Velli}, A.~{Yehle}, D.~{Werthimer}, and
  J.~R. {Wygant}.
\newblock {The FIELDS Instrument Suite for Solar Probe Plus. Measuring the
  Coronal Plasma and Magnetic Field, Plasma Waves and Turbulence, and Radio
  Signatures of Solar Transients}.
\newblock \emph{Space Science Rev.}, 204:\penalty0 49--82, December 2016.
\newblock \doi{10.1007/s11214-016-0244-5}.

\bibitem[Kasper et~al.(2016)Kasper, Abiad, Austin, Balat-Pichelin, Bale,
  Belcher, Berg, Bergner, Berthomier, Bookbinder, Brodu, Caldwell, Case,
  Chandran, Cheimets, Cirtain, Cranmer, Curtis, Daigneau, Dalton, Dasgupta,
  DeTomaso, Diaz-Aguado, Djordjevic, Donaskowski, Effinger, Florinski, Fox,
  Freeman, Gallagher, Gary, Gauron, Gates, Goldstein, Golub, Gordon, Gurnee,
  Guth, Halekas, Hatch, Heerikuisen, Ho, Hu, Johnson, Jordan, Korreck, Larson,
  Lazarus, Li, Livi, Ludlam, Maksimovic, McFadden, Marchant, Maruca, McComas,
  Messina, Mercer, Park, Peddie, Pogorelov, Reinhart, Richardson, Robinson,
  Rosen, Skoug, Slagle, Steinberg, Stevens, Szabo, Taylor, Tiu, Turin, Velli,
  Webb, Whittlesey, Wright, Wu, and Zank]{Kasper2016}
Justin~C. Kasper, Robert Abiad, Gerry Austin, Marianne Balat-Pichelin,
  Stuart~D. Bale, John~W. Belcher, Peter Berg, Henry Bergner, Matthieu
  Berthomier, Jay Bookbinder, Etienne Brodu, David Caldwell, Anthony~W. Case,
  Benjamin D.~G. Chandran, Peter Cheimets, Jonathan~W. Cirtain, Steven~R.
  Cranmer, David~W. Curtis, Peter Daigneau, Greg Dalton, Brahmananda Dasgupta,
  David DeTomaso, Millan Diaz-Aguado, Blagoje Djordjevic, Bill Donaskowski,
  Michael Effinger, Vladimir Florinski, Nichola Fox, Mark Freeman, Dennis
  Gallagher, S.~Peter Gary, Tom Gauron, Richard Gates, Melvin Goldstein, Leon
  Golub, Dorothy~A. Gordon, Reid Gurnee, Giora Guth, Jasper Halekas, Ken Hatch,
  Jacob Heerikuisen, George Ho, Qiang Hu, Greg Johnson, Steven~P. Jordan,
  Kelly~E. Korreck, Davin Larson, Alan~J. Lazarus, Gang Li, Roberto Livi,
  Michael Ludlam, Milan Maksimovic, James~P. McFadden, William Marchant,
  Bennet~A. Maruca, David~J. McComas, Luciana Messina, Tony Mercer, Sang Park,
  Andrew~M. Peddie, Nikolai Pogorelov, Matthew~J. Reinhart, John~D. Richardson,
  Miles Robinson, Irene Rosen, Ruth~M. Skoug, Amanda Slagle, John~T. Steinberg,
  Michael~L. Stevens, Adam Szabo, Ellen~R. Taylor, Chris Tiu, Paul Turin, Marco
  Velli, Gary Webb, Phyllis Whittlesey, Ken Wright, S.~T. Wu, and Gary Zank.
\newblock Solar wind electrons alphas and protons (sweap) investigation: Design
  of the solar wind and coronal plasma instrument suite for solar probe plus.
\newblock \emph{Space Science Reviews}, 204\penalty0 (1):\penalty0 131--186,
  Dec 2016.
\newblock ISSN 1572-9672.
\newblock \doi{10.1007/s11214-015-0206-3}.
\newblock URL \url{https://doi.org/10.1007/s11214-015-0206-3}.

\bibitem[{Verniero} et~al.(2020){Verniero}, {Larson}, {Livi}, {Rahmati},
  {McManus}, {Pyakurel}, {Klein}, {Bowen}, {Bonnell}, {Alterman}, {Whittlesey},
  {Malaspina}, {Bale}, {Kasper}, {Case}, {Goetz}, {Harvey}, {Korreck},
  {MacDowall}, {Pulupa}, {Stevens}, and {de Wit}]{Verniero2020}
J.~L. {Verniero}, D.~E. {Larson}, R.~{Livi}, A.~{Rahmati}, M.~D. {McManus},
  P.~Sharma {Pyakurel}, K.~G. {Klein}, T.~A. {Bowen}, J.~W. {Bonnell}, B.~L.
  {Alterman}, P.~L. {Whittlesey}, David~M. {Malaspina}, S.~D. {Bale}, J.~C.
  {Kasper}, A.~W. {Case}, K.~{Goetz}, P.~R. {Harvey}, K.~E. {Korreck}, R.~J.
  {MacDowall}, M.~{Pulupa}, M.~L. {Stevens}, and T.~Dudok {de Wit}.
\newblock {Parker Solar Probe Observations of Proton Beams Simultaneous with
  Ion-scale Waves}.
\newblock \emph{ApJS}, 248\penalty0 (1):\penalty0 5, May 2020.
\newblock \doi{10.3847/1538-4365/ab86af}.

\bibitem[{Klein} et~al.(2021){Klein}, {Verniero}, {Alterman}, {Bale}, {Case},
  {Kasper}, {Korreck}, {Larson}, {Lichko}, {Livi}, {McManus}, {Martinovi{\'c}},
  {Rahmati}, {Stevens}, and {Whittlesey}]{Klein2021}
K.~G. {Klein}, J.~L. {Verniero}, B.~{Alterman}, S.~{Bale}, A.~{Case}, J.~C.
  {Kasper}, K.~{Korreck}, D.~{Larson}, E.~{Lichko}, R.~{Livi}, M.~{McManus},
  M.~{Martinovi{\'c}}, A.~{Rahmati}, M.~{Stevens}, and P.~{Whittlesey}.
\newblock {Inferred Linear Stability of Parker Solar Probe Observations Using
  One- and Two-component Proton Distributions}.
\newblock \emph{ApJ}, 909\penalty0 (1):\penalty0 7, March 2021.
\newblock \doi{10.3847/1538-4357/abd7a0}.

\bibitem[{Verniero} et~al.(2022){Verniero}, {Chandran}, {Larson}, {Paulson},
  {Alterman}, {Badman}, {Bale}, {Bonnell}, {Bowen}, {de Wit}, {Kasper},
  {Klein}, {Lichko}, {Livi}, {McManus}, {Rahmati}, {Verscharen}, {Walters}, and
  {Whittlesey}]{Verniero2021}
J.~L. {Verniero}, B.~D.~G. {Chandran}, D.~E. {Larson}, K.~{Paulson}, B.~L.
  {Alterman}, S.~{Badman}, S.~D. {Bale}, J.~W. {Bonnell}, T.~A. {Bowen},
  T.~Dudok {de Wit}, J.~C. {Kasper}, K.~G. {Klein}, E.~{Lichko}, R.~{Livi},
  M.~D. {McManus}, A.~{Rahmati}, D.~{Verscharen}, J.~{Walters}, and P.~L.
  {Whittlesey}.
\newblock {Strong Perpendicular Velocity-space Diffusion in Proton Beams
  Observed by Parker Solar Probe}.
\newblock \emph{ApJ}, 924\penalty0 (2):\penalty0 112, January 2022.
\newblock \doi{10.3847/1538-4357/ac36d5}.

\bibitem[{Bowen} et~al.(2020{\natexlab{b}})]{Bowen2020b}
{{{T.~A.}}} {Bowen} et~al.
\newblock A merged search-coil and fluxgate magnetometer data product for
  parker solar probe fields.
\newblock \emph{JGR}, 125\penalty0 (5):\penalty0 e2020JA027813,
  2020{\natexlab{b}}.

\bibitem[{Dudok de Wit} et~al.(2022){Dudok de Wit}, {Krasnoselskikh},
  {Agapitov}, {Froment}, {Larosa}, {Bale}, {Bowen}, {Goetz}, {Harvey},
  {Jannet}, {Kretzschmar}, {MacDowall}, {Malaspina}, {Martin}, {Page},
  {Pulupa}, and {Revillet}]{DudokdeWit2022}
T.~{Dudok de Wit}, V.~V. {Krasnoselskikh}, O.~{Agapitov}, C.~{Froment},
  A.~{Larosa}, S.~D. {Bale}, T.~{Bowen}, K.~{Goetz}, P.~{Harvey}, G.~{Jannet},
  M.~{Kretzschmar}, R.~J. {MacDowall}, D.~{Malaspina}, P.~{Martin}, B.~{Page},
  M.~{Pulupa}, and C.~{Revillet}.
\newblock {First Results From the SCM Search-Coil Magnetometer on Parker Solar
  Probe}.
\newblock \emph{Journal of Geophysical Research (Space Physics)}, 127\penalty0
  (4):\penalty0 e30018, April 2022.
\newblock \doi{10.1029/2021JA030018}.

\bibitem[{Pulupa} et~al.(2017){Pulupa}, {Bale}, {Bonnell}, {Bowen}, {Carruth},
  {Goetz}, {Gordon}, {Harvey}, {Maksimovic}, {Mart{\'\i}nez-Oliveros},
  {Moncuquet}, {Saint-Hilaire}, {Seitz}, and {Sundkvist}]{Pulupa2017}
M.~{Pulupa}, S.~D. {Bale}, J.~W. {Bonnell}, T.~A. {Bowen}, N.~{Carruth},
  K.~{Goetz}, D.~{Gordon}, P.~R. {Harvey}, M.~{Maksimovic}, J.~C.
  {Mart{\'\i}nez-Oliveros}, M.~{Moncuquet}, P.~{Saint-Hilaire}, D.~{Seitz}, and
  D.~{Sundkvist}.
\newblock {The Solar Probe Plus Radio Frequency Spectrometer: Measurement
  requirements, analog design, and digital signal processing}.
\newblock \emph{Journal of Geophysical Research (Space Physics)}, 122\penalty0
  (3):\penalty0 2836--2854, March 2017.
\newblock \doi{10.1002/2016JA023345}.

\bibitem[{Sahraoui} et~al.(2009){Sahraoui}, {Goldstein}, {Robert}, and
  {Khotyaintsev}]{Sahraoui2009}
F.~{Sahraoui}, M.~L. {Goldstein}, P.~{Robert}, and Yu.~V. {Khotyaintsev}.
\newblock {Evidence of a Cascade and Dissipation of Solar-Wind Turbulence at
  the Electron Gyroscale}.
\newblock \emph{Physical Review Letters}, 102\penalty0 (23):\penalty0 231102,
  Jun 2009.
\newblock \doi{10.1103/PhysRevLett.102.231102}.

\bibitem[{Kiyani} et~al.(2009){Kiyani}, {Chapman}, {Khotyaintsev}, {Dunlop},
  and {Sahraoui}]{Kiyani2009}
K.~H. {Kiyani}, S.~C. {Chapman}, Yu.~V. {Khotyaintsev}, M.~W. {Dunlop}, and
  F.~{Sahraoui}.
\newblock {Global Scale-Invariant Dissipation in Collisionless Plasma
  Turbulence}.
\newblock \emph{Physical Review Letters}, 103\penalty0 (7):\penalty0 075006,
  Aug 2009.
\newblock \doi{10.1103/PhysRevLett.103.075006}.

\bibitem[{Bowen} et~al.(2020{\natexlab{c}}){Bowen}, {Mallet}, {Bale},
  {Bonnell}, {Case}, {Chandran}, {Chasapis}, {Chen}, {Duan}, {Dudok de Wit},
  {Goetz}, {Halekas}, {Harvey}, {Kasper}, {Korreck}, {Larson}, {Livi},
  {MacDowall}, {Malaspina}, {McManus}, {Pulupa}, {Stevens}, and
  {Whittlesey}]{Bowen2020c}
Trevor~A. {Bowen}, Alfred {Mallet}, Stuart~D. {Bale}, J.~W. {Bonnell},
  Anthony~W. {Case}, Benjamin D.~G. {Chandran}, Alexandros {Chasapis},
  Christopher H.~K. {Chen}, Die {Duan}, Thierry {Dudok de Wit}, Keith {Goetz},
  Jasper~S. {Halekas}, Peter~R. {Harvey}, J.~C. {Kasper}, Kelly~E. {Korreck},
  Davin {Larson}, Roberto {Livi}, Robert~J. {MacDowall}, David~M. {Malaspina},
  Michael~D. {McManus}, Marc {Pulupa}, Michael {Stevens}, and Phyllis
  {Whittlesey}.
\newblock {Constraining Ion-Scale Heating and Spectral Energy Transfer in
  Observations of Plasma Turbulence}.
\newblock \emph{PRL}, 125\penalty0 (2):\penalty0 025102, July
  2020{\natexlab{c}}.
\newblock \doi{10.1103/PhysRevLett.125.025102}.

\bibitem[{Torrence} and {Compo}(1998)]{TorrenceCompo1998}
Christopher {Torrence} and Gilbert~P. {Compo}.
\newblock {A Practical Guide to Wavelet Analysis.}
\newblock \emph{Bulletin of the American Meteorological Society}, 79\penalty0
  (1):\penalty0 61--78, January 1998.
\newblock
  \doi{10.1175/1520-0477(1998)079\textless{}0061:APGTWA\textgreater{}2.0.CO;2}.

\bibitem[{Howes} and {Quataert}(2010)]{HowesQuataert2010}
G.~G. {Howes} and E.~{Quataert}.
\newblock {On the Interpretation of Magnetic Helicity Signatures in the
  Dissipation Range Of Solar Wind Turbulence}.
\newblock \emph{The Astrophysical Journal Letters}, 709\penalty0 (1):\penalty0
  L49--L52, Jan 2010.
\newblock \doi{10.1088/2041-8205/709/1/L49}.

\bibitem[{Klein} et~al.(2014){Klein}, {Howes}, {TenBarge}, and
  {Podesta}]{Klein2014}
Kristopher~G. {Klein}, Gregory~G. {Howes}, Jason~M. {TenBarge}, and John~J.
  {Podesta}.
\newblock {Physical Interpretation of the Angle-dependent Magnetic Helicity
  Spectrum in the Solar Wind: The Nature of Turbulent Fluctuations near the
  Proton Gyroradius Scale}.
\newblock \emph{ApJ}, 785\penalty0 (2):\penalty0 138, April 2014.
\newblock \doi{10.1088/0004-637X/785/2/138}.

\bibitem[{Bowen} et~al.(2020{\natexlab{d}}){Bowen}, {Bale}, {Bonnell},
  {Larson}, {Mallet}, {McManus}, {Mozer}, {Pulupa}, {Vasko}, {Verniero},
  {Psp/Fields Team}, and {Psp/Sweap Teams}]{Bowen2020d}
Trevor~A. {Bowen}, Stuart~D. {Bale}, J.~W. {Bonnell}, Davin {Larson}, Alfred
  {Mallet}, Michael~D. {McManus}, Forrest~S. {Mozer}, Marc {Pulupa}, Ivan~Y.
  {Vasko}, J.~L. {Verniero}, {Psp/Fields Team}, and {Psp/Sweap Teams}.
\newblock {The Electromagnetic Signature of Outward Propagating Ion-scale
  Waves}.
\newblock \emph{ApJ}, 899\penalty0 (1):\penalty0 74, August 2020{\natexlab{d}}.
\newblock \doi{10.3847/1538-4357/ab9f37}.

\bibitem[{Fredricks} and {Coroniti}(1976)]{Fredricks1976}
R.~W. {Fredricks} and F.~V. {Coroniti}.
\newblock {Ambiguities in the deduction of rest frame fluctuation spectrums
  from spectrums computed in moving frames}.
\newblock \emph{JGR}, 81\penalty0 (A31):\penalty0 5591--5595, November 1976.
\newblock \doi{10.1029/JA081i031p05591}.

\bibitem[{Horbury} et~al.(2008){Horbury}, {Forman}, and {Oughton}]{Horbury2008}
Timothy~S. {Horbury}, Miriam {Forman}, and Sean {Oughton}.
\newblock {Anisotropic Scaling of Magnetohydrodynamic Turbulence}.
\newblock \emph{\prl}, 101\penalty0 (17):\penalty0 175005, October 2008.
\newblock \doi{10.1103/PhysRevLett.101.175005}.

\bibitem[{Horbury} et~al.(2012){Horbury}, {Wicks}, and {Chen}]{Horbury2012}
T.~S. {Horbury}, R.~T. {Wicks}, and C.~H.~K. {Chen}.
\newblock {Anisotropy in Space Plasma Turbulence: Solar Wind Observations}.
\newblock \emph{Space Sci. Rev.}, 172\penalty0 (1-4):\penalty0 325--342,
  November 2012.
\newblock \doi{10.1007/s11214-011-9821-9}.

\bibitem[{Dum} et~al.(1980){Dum}, {Marsch}, and {Pilipp}]{Dum1980}
C.~T. {Dum}, E.~{Marsch}, and W.~{Pilipp}.
\newblock {Determination of wave growth from measured distribution functions
  and transport theory}.
\newblock \emph{Journal of Plasma Physics}, 23\penalty0 (1):\penalty0 91--113,
  February 1980.
\newblock \doi{10.1017/S0022377800022170}.

\bibitem[{Vi{\~n}as} and {Gurgiolo}(2009)]{Vinas2009}
Adolfo~F. {Vi{\~n}as} and Chris {Gurgiolo}.
\newblock {Spherical harmonic analysis of particle velocity distribution
  function: Comparison of moments and anisotropies using Cluster data}.
\newblock \emph{Journal of Geophysical Research (Space Physics)}, 114\penalty0
  (A1):\penalty0 A01105, January 2009.
\newblock \doi{10.1029/2008JA013633}.

\bibitem[{Servidio} et~al.(2017){Servidio}, {Chasapis}, {Matthaeus}, {Perrone},
  {Valentini}, {Parashar}, {Veltri}, {Gershman}, {Russell}, {Giles},
  {Fuselier}, {Phan}, and {Burch}]{Servidio2017}
S.~{Servidio}, A.~{Chasapis}, W.~H. {Matthaeus}, D.~{Perrone}, F.~{Valentini},
  T.~N. {Parashar}, P.~{Veltri}, D.~{Gershman}, C.~T. {Russell}, B.~{Giles},
  S.~A. {Fuselier}, T.~D. {Phan}, and J.~{Burch}.
\newblock {Magnetospheric Multiscale Observation of Plasma Velocity-Space
  Cascade: Hermite Representation and Theory}.
\newblock \emph{\prl}, 119\penalty0 (20):\penalty0 205101, November 2017.
\newblock \doi{10.1103/PhysRevLett.119.205101}.

\bibitem[{Parker} and {Dellar}(2015)]{Parker2015}
Joseph~T. {Parker} and Paul~J. {Dellar}.
\newblock {Fourier-Hermite spectral representation for the Vlasov-Poisson
  system in the weakly collisional limit}.
\newblock \emph{Journal of Plasma Physics}, 81\penalty0 (2):\penalty0
  305810203, April 2015.
\newblock \doi{10.1017/S0022377814001287}.

\bibitem[{Schekochihin} et~al.(2016){Schekochihin}, {Parker}, {Highcock},
  {Dellar}, {Dorland}, and {Hammett}]{Schekochihin2016}
A.~A. {Schekochihin}, J.~T. {Parker}, E.~G. {Highcock}, P.~J. {Dellar},
  W.~{Dorland}, and G.~W. {Hammett}.
\newblock {Phase mixing versus nonlinear advection in drift-kinetic plasma
  turbulence}.
\newblock \emph{Journal of Plasma Physics}, 82\penalty0 (2):\penalty0
  905820212, April 2016.
\newblock \doi{10.1017/S0022377816000374}.

\bibitem[{Squire} et~al.(2020){Squire}, {Chandran}, and {Meyrand}]{Squire2020}
J.~{Squire}, B.~D.~G. {Chandran}, and R.~{Meyrand}.
\newblock {In-situ Switchback Formation in the Expanding Solar Wind}.
\newblock \emph{ApJL}, 891\penalty0 (1):\penalty0 L2, March 2020.
\newblock \doi{10.3847/2041-8213/ab74e1}.

\bibitem[{Chandran} and {Perez}(2019)]{Chandran2019}
Benjamin D.~G. {Chandran} and Jean~C. {Perez}.
\newblock {Reflection-driven magnetohydrodynamic turbulence in the solar
  atmosphere and solar wind}.
\newblock \emph{Journal of Plasma Physics}, 85\penalty0 (4):\penalty0
  905850409, Aug 2019.
\newblock \doi{10.1017/S0022377819000540}.

\bibitem[{Hellinger} et~al.(2013){Hellinger}, {Tr{\'a}Vn{\'\i}{\v{c}}ek},
  {{\v{S}}tver{\'a}k}, {Matteini}, and {Velli}]{Hellinger2013}
Petr {Hellinger}, Pavel~M. {Tr{\'a}Vn{\'\i}{\v{c}}ek}, {\v{S}}t{\v{e}}p{\'a}n
  {{\v{S}}tver{\'a}k}, Lorenzo {Matteini}, and Marco {Velli}.
\newblock {Proton thermal energetics in the solar wind: Helios reloaded}.
\newblock \emph{Journal of Geophysical Research (Space Physics)}, 118\penalty0
  (4):\penalty0 1351--1365, April 2013.
\newblock \doi{10.1002/jgra.50107}.

\bibitem[{Sasikumar Raja} et~al.(2021){Sasikumar Raja}, {Subramanian},
  {Ingale}, {Ramesh}, and {Maksimovic}]{Raja2021}
K.~{Sasikumar Raja}, Prasad {Subramanian}, Madhusudan {Ingale}, R.~{Ramesh},
  and Milan {Maksimovic}.
\newblock {Turbulent Proton Heating Rate in the Solar Wind from 5-45
  R$_{{\ensuremath{\odot}}}$}.
\newblock \emph{\apj}, 914\penalty0 (2):\penalty0 137, June 2021.
\newblock \doi{10.3847/1538-4357/abfcd1}.

\bibitem[{Politano} and {Pouquet}(1998)]{Politano1998a}
H.~{Politano} and A.~{Pouquet}.
\newblock {von K{\'a}rm{\'a}n-Howarth equation for magnetohydrodynamics and its
  consequences on third-order longitudinal structure and correlation
  functions}.
\newblock \emph{Phys Rev E.}, 57\penalty0 (1):\penalty0 R21--R24, January 1998.
\newblock \doi{10.1103/PhysRevE.57.R21}.

\bibitem[{Bandyopadhyay} et~al.(2020){Bandyopadhyay}, {Goldstein}, {Maruca},
  {Matthaeus}, {Parashar}, {Ruffolo}, {Chhiber}, {Usmanov}, {Chasapis},
  {Qudsi}, {Bale}, {Bonnell}, {Dudok de Wit}, {Goetz}, {Harvey}, {MacDowall},
  {Malaspina}, {Pulupa}, {Kasper}, {Korreck}, {Case}, {Stevens}, {Whittlesey},
  {Larson}, {Livi}, {Klein}, {Velli}, and {Raouafi}]{Bandyopadhyay2020}
Riddhi {Bandyopadhyay}, M.~L. {Goldstein}, B.~A. {Maruca}, W.~H. {Matthaeus},
  T.~N. {Parashar}, D.~{Ruffolo}, R.~{Chhiber}, A.~{Usmanov}, A.~{Chasapis},
  R.~{Qudsi}, Stuart~D. {Bale}, J.~W. {Bonnell}, Thierry {Dudok de Wit}, Keith
  {Goetz}, Peter~R. {Harvey}, Robert~J. {MacDowall}, David~M. {Malaspina}, Marc
  {Pulupa}, J.~C. {Kasper}, K.~E. {Korreck}, A.~W. {Case}, M.~{Stevens},
  P.~{Whittlesey}, D.~{Larson}, R.~{Livi}, K.~G. {Klein}, M.~{Velli}, and
  N.~{Raouafi}.
\newblock {Enhanced Energy Transfer Rate in Solar Wind Turbulence Observed near
  the Sun from Parker Solar Probe}.
\newblock \emph{ApJs}, 246\penalty0 (2):\penalty0 48, February 2020.
\newblock \doi{10.3847/1538-4365/ab5dae}.

\bibitem[{Andr{\'e}s} et~al.(2022){Andr{\'e}s}, {Sahraoui}, {Huang}, {Hadid},
  and {Galtier}]{Andres2022}
N.~{Andr{\'e}s}, F.~{Sahraoui}, S.~{Huang}, L.~Z. {Hadid}, and S.~{Galtier}.
\newblock {The incompressible energy cascade rate in anisotropic solar wind
  turbulence}.
\newblock \emph{A\&A}, 661:\penalty0 A116, May 2022.
\newblock \doi{10.1051/0004-6361/202142994}.

\bibitem[{Stawarz} et~al.(2009){Stawarz}, {Smith}, {Vasquez}, {Forman}, and
  {MacBride}]{Stawarz2009}
Joshua~E. {Stawarz}, Charles~W. {Smith}, Bernard~J. {Vasquez}, Miriam~A.
  {Forman}, and Benjamin~T. {MacBride}.
\newblock {The Turbulent Cascade and Proton Heating in the Solar Wind at 1 AU}.
\newblock \emph{ApJ}, 697\penalty0 (2):\penalty0 1119--1127, June 2009.
\newblock \doi{10.1088/0004-637X/697/2/1119}.

\bibitem[{Osman} et~al.(2011){Osman}, {Wan}, {Matthaeus}, {Weygand}, and
  {Dasso}]{Osman2011PRL}
K.~T. {Osman}, M.~{Wan}, W.~H. {Matthaeus}, J.~M. {Weygand}, and S.~{Dasso}.
\newblock {Anisotropic Third-Moment Estimates of the Energy Cascade in Solar
  Wind Turbulence Using Multispacecraft Data}.
\newblock \emph{\prl}, 107\penalty0 (16):\penalty0 165001, October 2011.
\newblock \doi{10.1103/PhysRevLett.107.165001}.

\bibitem[{Chandran} et~al.(2010{\natexlab{c}}){Chandran}, {Li}, {Rogers},
  {Quataert}, and {Germaschewski}]{Chandran2010}
Benjamin D.~G. {Chandran}, Bo~{Li}, Barrett~N. {Rogers}, Eliot {Quataert}, and
  Kai {Germaschewski}.
\newblock {Perpendicular Ion Heating by Low-frequency Alfv{\'e}n-wave
  Turbulence in the Solar Wind}.
\newblock \emph{The Astrophysical Journal}, 720\penalty0 (1):\penalty0
  503--515, Sep 2010{\natexlab{c}}.
\newblock \doi{10.1088/0004-637X/720/1/503}.

\bibitem[{Squire} et~al.(2022){Squire}, {Meyrand}, {Kunz}, {Arzamasskiy},
  {Schekochihin}, and {Quataert}]{Squire2022}
Jonathan {Squire}, Romain {Meyrand}, Matthew~W. {Kunz}, Lev {Arzamasskiy},
  Alexander~A. {Schekochihin}, and Eliot {Quataert}.
\newblock {High-frequency heating of the solar wind triggered by low-frequency
  turbulence}.
\newblock \emph{Nature Astronomy}, 6:\penalty0 715--723, March 2022.
\newblock \doi{10.1038/s41550-022-01624-z}.

\bibitem[{Verscharen} et~al.(2013){Verscharen}, {Bourouaine}, and
  {Chandran}]{Verscharen2013}
Daniel {Verscharen}, Sofiane {Bourouaine}, and Benjamin D.~G. {Chandran}.
\newblock {Instabilities Driven by the Drift and Temperature Anisotropy of
  Alpha Particles in the Solar Wind}.
\newblock \emph{ApJ}, 773\penalty0 (2):\penalty0 163, August 2013.
\newblock \doi{10.1088/0004-637X/773/2/163}.

\bibitem[{Shebalin} et~al.(1983){Shebalin}, {Matthaeus}, and
  {Montgomery}]{Shebalin1983}
J.~V. {Shebalin}, W.~H. {Matthaeus}, and D.~{Montgomery}.
\newblock {Anisotropy in MHD turbulence due to a mean magnetic field}.
\newblock \emph{Journal of Plasma Physics}, 29\penalty0 (3):\penalty0 525--547,
  June 1983.
\newblock \doi{10.1017/S0022377800000933}.

\bibitem[{Goldreich} and {Sridhar}(1995)]{GS95}
P.~{Goldreich} and S.~{Sridhar}.
\newblock {Toward a Theory of Interstellar Turbulence. II. Strong Alfvenic
  Turbulence}.
\newblock \emph{The Astrophysical Journal}, 438:\penalty0 763, Jan 1995.
\newblock \doi{10.1086/175121}.

\bibitem[{Meyrand} et~al.(2021){Meyrand}, {Squire}, {Schekochihin}, and
  {Dorland}]{Meyrand2021}
R.~{Meyrand}, J.~{Squire}, A.~A. {Schekochihin}, and W.~{Dorland}.
\newblock {On the violation of the zeroth law of turbulence in space plasmas}.
\newblock \emph{Journal of Plasma Physics}, 87\penalty0 (3):\penalty0
  535870301, May 2021.
\newblock \doi{10.1017/S0022377821000489}.

\bibitem[{Stawarz} et~al.(2010){Stawarz}, {Smith}, {Vasquez}, {Forman}, and
  {MacBride}]{Stawarz2010}
Joshua~E. {Stawarz}, Charles~W. {Smith}, Bernard~J. {Vasquez}, Miriam~A.
  {Forman}, and Benjamin~T. {MacBride}.
\newblock {The Turbulent Cascade for High Cross-Helicity States at 1 AU}.
\newblock \emph{ApJ}, 713\penalty0 (2):\penalty0 920--934, April 2010.
\newblock \doi{10.1088/0004-637X/713/2/920}.

\bibitem[{Chen} et~al.(2020){Chen}, {Bale}, {Bonnell}, {Borovikov}, {Bowen},
  {Burgess}, {Case}, {Chandran}, {de Wit}, {Goetz}, {Harvey}, {Kasper},
  {Klein}, {Korreck}, {Larson}, {Livi}, {MacDowall}, {Malaspina}, {Mallet},
  {McManus}, {Moncuquet}, {Pulupa}, {Stevens}, and {Whittlesey}]{Chen2020}
C.~H.~K. {Chen}, S.~D. {Bale}, J.~W. {Bonnell}, D.~{Borovikov}, T.~A. {Bowen},
  D.~{Burgess}, A.~W. {Case}, B.~D.~G. {Chandran}, T.~Dudok {de Wit},
  K.~{Goetz}, P.~R. {Harvey}, J.~C. {Kasper}, K.~G. {Klein}, K.~E. {Korreck},
  D.~{Larson}, R.~{Livi}, R.~J. {MacDowall}, D.~M. {Malaspina}, A.~{Mallet},
  M.~D. {McManus}, M.~{Moncuquet}, M.~{Pulupa}, M.~L. {Stevens}, and
  P.~{Whittlesey}.
\newblock {The Evolution and Role of Solar Wind Turbulence in the Inner
  Heliosphere}.
\newblock \emph{ApJS}, 246\penalty0 (2):\penalty0 53, February 2020.
\newblock \doi{10.3847/1538-4365/ab60a3}.

\bibitem[{Duan} et~al.(2020){Duan}, {Bowen}, {Chen}, {Mallet}, {He}, {Bale},
  {Vech}, {Kasper}, {Pulupa}, {Bonnell}, {Case}, {de Wit}, {Goetz}, {Harvey},
  {Korreck}, {Larson}, {Livi}, {MacDowall}, {Malaspina}, {Stevens}, and
  {Whittlesey}]{Duan2020}
Die {Duan}, Trevor~A. {Bowen}, Christopher H.~K. {Chen}, Alfred {Mallet},
  Jiansen {He}, Stuart~D. {Bale}, Daniel {Vech}, J.~C. {Kasper}, Marc {Pulupa},
  John~W. {Bonnell}, Anthony~W. {Case}, Thierry~Dudok {de Wit}, Keith {Goetz},
  Peter~R. {Harvey}, Kelly~E. {Korreck}, Davin {Larson}, Roberto {Livi},
  Robert~J. {MacDowall}, David~M. {Malaspina}, Michael {Stevens}, and Phyllis
  {Whittlesey}.
\newblock {The Radial Dependence of Proton-scale Magnetic Spectral Break in
  Slow Solar Wind during PSP Encounter 2}.
\newblock \emph{ApJS}, 246\penalty0 (2):\penalty0 55, February 2020.
\newblock \doi{10.3847/1538-4365/ab672d}.

\bibitem[{D} et~al.(2021){D}, {He}, {Bowen}, {Woodham}, {Wang}, {Chen},
  {Mallet}, and {Bale}]{Duan2021}
Die {D}, Jiansen {He}, Trevor~A. {Bowen}, Lloyd~D. {Woodham}, Tieyan {Wang},
  Christopher H.~K. {Chen}, Alfred {Mallet}, and Stuart~D. {Bale}.
\newblock {Anisotropy of Solar Wind Turbulence in the Inner Heliosphere at
  Kinetic Scales: PSP Observations}.
\newblock \emph{ApJL}, 915\penalty0 (1):\penalty0 L8, July 2021.
\newblock \doi{10.3847/2041-8213/ac07ac}.

\bibitem[{Mozer} et~al.(2020){Mozer}, {Agapitov}, {Bale}, {Bonnell}, {Bowen},
  and {Vasko}]{Mozer2020}
F.~S. {Mozer}, O.~V. {Agapitov}, S.~D. {Bale}, J.~W. {Bonnell}, T.~A. {Bowen},
  and I.~{Vasko}.
\newblock {DC and Low-Frequency Electric Field Measurements on the Parker Solar
  Probe}.
\newblock \emph{Journal of Geophysical Research (Space Physics)}, 125\penalty0
  (9):\penalty0 e27980, September 2020.
\newblock \doi{10.1029/2020JA027980}.

\bibitem[{McManus} et~al.(2020){McManus}, {Bowen}, {Mallet}, {Chen},
  {Chandran}, {Bale}, {Larson}, {Dudok de Wit}, {Kasper}, {Stevens},
  {Whittlesey}, {Livi}, {Korreck}, {Goetz}, {Harvey}, {Pulupa}, {MacDowall},
  {Malaspina}, {Case}, and {Bonnell}]{McManus2020}
Michael~D. {McManus}, Trevor~A. {Bowen}, Alfred {Mallet}, Christopher H.~K.
  {Chen}, Benjamin D.~G. {Chandran}, Stuart~D. {Bale}, Davin~E. {Larson},
  Thierry {Dudok de Wit}, J.~C. {Kasper}, Michael {Stevens}, Phyllis
  {Whittlesey}, Roberto {Livi}, Kelly~E. {Korreck}, Keith {Goetz}, Peter~R.
  {Harvey}, Marc {Pulupa}, Robert~J. {MacDowall}, David~M. {Malaspina},
  Anthony~W. {Case}, and J.~W. {Bonnell}.
\newblock {Cross Helicity Reversals in Magnetic Switchbacks}.
\newblock \emph{ApJS}, 246\penalty0 (2):\penalty0 67, February 2020.
\newblock \doi{10.3847/1538-4365/ab6dce}.

\end{thebibliography}
\end{document}